\documentclass[Afour,sageh,times]{sagej}
\usepackage{mathptmx}
\usepackage{moreverb,url}

\usepackage[colorlinks,bookmarksopen,bookmarksnumbered,citecolor=red,urlcolor=red]{hyperref}

\usepackage{listings}
\usepackage[export]{adjustbox}
\usepackage{xcolor}

\usepackage{array}
\newcolumntype{L}[1]{>{\raggedright}m{#1}}

\newcommand\BibTeX{{\rmfamily B\kern-.05em \textsc{i\kern-.025em b}\kern-.08em
T\kern-.1667em\lower.7ex\hbox{E}\kern-.125emX}}

\newcommand{\bq}{\begin{equation}}
\newcommand{\eq}{\end{equation}}

\newcommand{\byte}{\mbox{B}}
\newcommand{\second}{\mbox{s}}

\newcommand{\flop}{\mbox{flop}}

\newcommand{\cycle}{\mbox{cy}}
\newcommand{\cycles}{\mbox{cy}}

\newcommand{\TBS}{\mbox{T\byte/\second}}

\newcommand{\GFS}{\mbox{G\flop/\second}}

\newcommand{\GHZ}{\mbox{GHz}}

\newcommand{\KB}{\mbox{kB}}
\newcommand{\MB}{\mbox{MB}}

\newcommand{\eos}{\;.}
\newcommand{\cma}{\;,}

\newcommand{\Rlm}{Roof\-line model}

\newcommand{\olsep}{\|}
\newcommand{\nolsep}{|}
\newcommand{\ecmspace}{\,}
\newcommand{\ecm}[6]{\mbox{$\left\{{#1}\ecmspace\olsep\ecmspace {#2}\ecmspace\nolsep\ecmspace {#3}\ecmspace\nolsep\ecmspace {#4}\ecmspace\nolsep\ecmspace {#5}\right\}\ecmspace{#6}$}}
\newcommand{\epsep}{\rceil}
\newcommand{\ecmp}[5]{\mbox{$\left\{{#1}\ecmspace\epsep\ecmspace {#2}\ecmspace\epsep\ecmspace {#3}\ecmspace\epsep\ecmspace {#4}\right\}\ecmspace{#5}$}}
\newcommand{\ecme}[5]{\mbox{$\left({#1}\ecmspace\epsep\ecmspace {#2}\ecmspace\epsep\ecmspace {#3}\ecmspace\epsep\ecmspace {#4}\right)\ecmspace{#5}$}}

\begin{document}

\runninghead{Cremonesi et al.}

\title{Analytic Performance Modeling and Analysis of Detailed Neuron Simulations}

\author{Francesco Cremonesi\affilnum{1}, Georg Hager\affilnum{2}, Gerhard Wellein\affilnum{3} and Felix Sch\"urmann\affilnum{1}}

\affiliation{%
    \affilnum{1}Blue Brain Project,
    Brain Mind Institute,
    \'Ecole polytechnique f\'ed\'erale de Lausanne\\
    \affilnum{2}Erlangen Regional Computing Center,
    Friedrich-Alexander Universit\"at Erlangen-N\"urnberg\\
    \affilnum{3}Department of Computer Science,
    Friedrich-Alexander Universit\"at Erlangen-N\"urnberg}

\corrauth{%
    Felix Sch\"urman,
    Campus Biotech,
    B1 4 282.040
    Ch. des Mines 9,
    CH-1202 Gen\`eve }

\email{felix.schuermann@epfl.ch}

\begin{abstract}
  Big science initiatives are trying to reconstruct and model the brain by attempting to simulate brain tissue at larger scales and with increasingly more biological detail than previously thought possible. The exponential growth of parallel computer performance has been supporting these developments, and at the same time maintainers of neuroscientific simulation code have strived to optimally and efficiently exploit new hardware features. Current state of the art software for the simulation of biological networks has so far been developed using performance engineering practices, but a thorough analysis and modeling of the computational and performance characteristics, especially in the case of morphologically detailed neuron simulations, is lacking. Other computational sciences have successfully used analytic performance engineering and modeling methods to gain insight on the computational properties of simulation kernels, aid developers in performance optimizations and eventually drive co-design efforts, but to our knowledge a model-based performance analysis of neuron simulations has not yet been conducted.

  We present a detailed study of the shared-memory performance of morphologically detailed neuron simulations based on the Execution-Cache-Memory (ECM) performance model. We demonstrate that this model can deliver accurate predictions of the runtime of almost all the kernels that constitute the neuron models under investigation. The gained insight is used to identify the main governing mechanisms underlying performance bottlenecks in the simulation. The implications of this analysis on the optimization of neural simulation software and eventually co-design of future hardware architectures are discussed. In this sense, our work represents a valuable conceptual and quantitative contribution to understanding the performance properties of biological networks simulations.
  \end{abstract}

\keywords{Analytic performance modeling, execution-cache-memory model, biological neural networks, morphologically detailed neuron models}

\maketitle

\section{Introduction and related work}

\subsection{Neuron simulations}
Understanding the biological and theoretical principles underlying the brain's physiological and cognitive functions is a great challenge for modern science.
Exploiting the greater availability of data and resources, new computational approaches based on mathematical modeling and simulations have been developed to bridge the gap between the observed structural and functional complexity of the brain and the rather sparse experimental data, such as the works of~\cite{izhikevich2008large,potjans2012cell,markram2015reconstruction} and~\cite{schuecker2017fundamental}.

Simulations of biological neurons are characterized by demanding performance and memory requirements: a neuron can have up to 10,000 connections and must track separate states and events for each one; the model for a single neuron can be very detailed itself and contain up to 20,000 differential equations; neurons are very dense, and a small piece of tissue of roughly 1~mm$^3$ can contain up to 100,000 neurons; finally, very fast axonal connections and current transients can limit the simulation timestep to 0.1~ms or even lower.
Therefore, developers have gone to great lengths optimizing the memory requirements of the connectivity infrastructure in~\cite{jordan2018extremely}, the efficiency of the parallel communication algorithm in~\cite{hines2011comparison} and~\cite{ananthanarayanan2007anatomy}, the scalability of data distribution in~\cite{kozloski2011ultrascalable} and even the parallel assembly of the neural network in~\cite{ippen2017constructing}.
While these efforts improve the performance of the distributed simulations, little is still known about the intrinsic single-core and shared memory performance properties of neuron simulations.
On the other hand, the work of~\cite{zenke2014limits} studied the performance of shared-memory simulations of biological neurons.
However their analysis is mainly based on empirical performance analysis and is centered on current-based point neuron simulations, a formalism that discards information about a neuron's arborization.

The assumptions underlying brain simulations are very diverse, leading to a wide range of models across several orders of magnitude of spatial and time scales and thus to a complex landscape of simulation strategies, as summarized in the reviews by~\cite{brette2007simulation} and~\cite{tikidji2017software}.
In this work we focus on the simulation of morphologically detailed neurons based on the popular neuroscientific software NEURON presented in~\cite{carnevale2006neuron}, which implements a modeling paradigm that includes details about a neuron's individual morphology as well as its connectivity and allows to easily introduce custom models in the system.
Our purpose is to extract the fundamental computational properties of the simulations of detailed biological networks and understand their relationship with modern microprocessor architectures.

\subsection{The need for analytic performance modeling}

An analytic performance model is a simplified description of
the interactions between software and hardware together with a recipe
for generating  predictions of execution time.
Such a model must be simple to be tractable but also elaborate
enough to produce useful predictions.

Purely analytic (a.k.a.\ \emph{first-principles} or \emph{white-box})
models are based on known technical details of the hardware and some
assumptions about how the software executes. The textbook example
of a white-box model is the \Rlm\ by \cite{roofline:2009} for loop
performance prediction. The accuracy of such
predictions depends crucially on the reliability of low-level details. A
lack of predictive power challenges the underlying assumptions and,
once corrected, often leads to better insight. 

\emph{Black-box} models, on the other hand, are ideally unaware of
code and hardware specifics; measured data is used to identify crucial
influence factors for the metrics to be
modeled (see. e.g., \cite{calotoiu_ea:2013:modeling}).  One can then predict
properties of arbitrary code, or play with parameters to explore
design spaces. Black-box models have a wider range of applicability:
Even if low-level hardware information is lacking they still provide
predictive power. Wrong predictions, however, may be rooted in
inappropriate fitting procedures and do not directly lead to better
insight.

In this work we choose the analytic approach combined with some
phenomenological input, which makes the model a \emph{gray-box} model.
Analytic modeling has several decisive advantages that make it more
suitable for delivering the insight we are looking for. First, it
allows for \emph{universality identification}, which means that
some behavior in hardware-software interaction is valid for a wide
range of microarchitectures of some kind.
Second, it enables the \emph{identification of governing mechanisms}: Since the model
pinpoints the actual performance bottlenecks in the hardware, classes
of codes with similar behavior are readily identified. This insight
directly leads to possible co-design approaches. And third,
analytic models provide \emph{insight via model failure}, as described above.

\subsection{The ECM performance model for multicore processors}

The Execution-Cache-Memory (ECM) model takes into account predictions
of single-threaded in-core execution time and data transfers through
the complete cache hierarchy for steady-state loops. These predictions
can be put together in different ways, depending on the CPU
architecture. One starts with ``optimistic'' transfer times through
the memory hierarchy:
\bq
T_i=\frac{V_i}{b_i}\cma
\eq
where $b_i$ is the bandwidth of data path $i$ and $V_i$ is the
data volume transferred over it for some definite unit of work
(e.g., one iteration of a loop). For convenience we use
a compact notation for such predictions, e.g.:
\bq\label{eq_ecm_contrib}
\{T_{L1L2}\ecmspace\nolsep\ecmspace T_{L2L3}\ecmspace\nolsep\ecmspace T_{L3Mem}\}=
\{4\ecmspace\nolsep\ecmspace 8\ecmspace\nolsep\ecmspace 18.4\}\,\text{cy/it}\eos
\eq
While the $b_i$ are machine properties, the $V_i$ must be
obtained by analysis, i.e., using knowledge about how data flows
through the system (see, e.g.,~\cite{hager2018applying}).

As for code execution in the core, one has to set up a model for
the execution time of the loop, which may be as simple as assuming
maximum throughput for all instructions, or as complex as
considering the full critical path. Tools such as IACA,
the Architecture Code Analyzer by~\cite{IACA} can help with this task. 

On all recent Intel server microarchitectures it turns out that the
single-core model yields the best predictions if one assumes no (temporal) overlap
of data transfers through the cache hierarchy and between the L1 cache
and registers, while in-core execution (such as arithmetic) shows full
overlap. For a dataset with a considerable amount of memory transfers,
the model thus predicts:
\bq
T_{ECM}^{Mem} = \max\left(T_{OL}, T_{nOL}
+T_{L1L2}+T_{L2L3}+T_{L3Mem}\right)\eos
\eq
Here, $T_{OL}$ is the part of the in-core execution that is
unrelated to data transfers, such as arithmetic, while $T_{nOL}$
is the time (cycles) required to retire load instructions.
For runtime predictions we use the following shorthand notation, to be distinguished from (\ref{eq_ecm_contrib}) by the use of $\rceil$ as delimiter:
\bq
\ecmp{T_{ECM}^{L1}}{T_{ECM}^{L2}}{T_{ECM}^{L3}}{T_{ECM}^{Mem}}{\text{cy/it}},
\eq
where $T_{ECM}^{X}$ denotes the runtime prediction if data comes from the $X^{th}$ level of the memory hierarchy. For presenting measurements we substitute the curly braces by
parentheses or omit them altogether.

Scalability is assumed to be perfect until a bandwidth
bottleneck is hit. Since the memory interface is the only multi-core
bandwidth bottleneck on Intel processors, the predicted execution time
is for $n$ cores is 
\bq\label{eq_T_ECM}
T_{ECM}=\max\left(\frac{T_{ECM}^{Mem}}{n},T_{L3Mem}\right)\eos
\eq
The bandwidth saturation point, i.e., the number of cores
required for saturation, is readily obtained from this
expression:
\bq
n_ S=\left\lceil \frac{T_{ECM}^{Mem}}{T_{L3Mem}} \right\rceil
\eq

A full account of the ECM model would exceed the
scope of this paper, so we refer to \cite{sthw15} and \cite{Hofmann:2017} for a recent
discussion.

\subsection{Contributions and organization of this paper}

In this work we make the following contributions:
\begin{itemize}
\item We demonstrate that the analytic ECM performance model can be applied
  successfully to nontrivial loop kernels with a wide range of
  different performance features. Although there are considerable
  error margins in some cases, a very good qualitative understanding
  can be achieved.  We also identify cases where the model needs
  corrections or cannot be applied sensibly: Strong latency
  components in the data transfers and long critical paths in the core
  execution. While the former is beyond the applicability
  of the model in its current form, the latter does not hinder
  the derivation of useful conclusions.
\item We apply the ECM model for the first time to the Intel Skylake-X
  processor architecture, whose cache hierarchy is different from
  earlier Intel designs.
\item We give clear guidelines for co-designing an ``ideal'' processor
  architecture for neuron simulations. In particular, we spot wide
  SIMD capabilities as a crucial ingredient in achieving memory
  bandwidth saturation. A low core count part with a high clock speed
  and wide SIMD units (such as AVX-512) will present the most
  cost-effective hardware platform. Cache size is inconsequential for
  most kernels.
\item As a consequence we can also predict if linear algebra
  and spike delivery kernels may become bottlenecks if
  a very large degree of parallelism could be exposed,
  and what hardware features would be required to avoid this.
\end{itemize}
This paper is organized as follows: In Section~\ref{sec:environ} we
give details on the software and hardware environment under
investigation, including preliminary performance observations. In
Section~\ref{sec:modeling} we construct and validate ECM performance
models for the important kernel classes in NEURON\@. In Section~\ref{sec:discussion}
we summarize and discuss the findings in order to pinpoint
the pivotal components of processor architectures in terms
of neuron simulation performance, and give an outlook to future work.

We provide a reproducibility appendix as a downloadable release file
at~\cite{BBP-ECM-RA}, which should enable the
interested reader to re-run our experiments and reproduce the relevant
performance data.

\section{Application and simulation environment}\label{sec:environ}
\subsection{Target architectures and programming environment}
We apply the ECM model introduced by~\cite{treibig2010introducing} and refined by~\cite{sthw15} on two Intel processors with different micro-architectures: the Ivy Bridge (IVB) Intel(R) Xeon(R) E5-2660v2 and the Skylake (SKX) Intel(R) Xeon(R) Gold 6140 (with Sub-NUMA clustering turned off).
The ECM model for the IVB architecture has been extensively studied
by~\cite{Hofmann:2017} and~\cite{hammer2017kerncraft}.
\begin{table}
\small\sf\centering
\caption{Hardware characteristics of the
  target CPU architectures.\label{tab_hw}}
\begin{tabular}{lcc}
    \toprule
    Characteristic & IVB & SKX \\
    \midrule
    CPU freq [GHz] & $2.2$ & $2.3$ \\
    Uncore freq [GHz] & $2.2$ & $2.3$ \\
    Mem BW (meas.) [GB/s]  & $40$  & $105$ \\
    LD/ST throughput per \cycle: & & \\
    ~~~AVX(2), AVX512 & 1 LD, $\frac{1}{2}$ST & 2 LD, 1 ST \\
    ~~~SSE, scalar & 2 LD $||$ 1 LD, 1 ST & 2 LD, 1 ST\\
    AGUs  & 2 & 2 + 1 simple \\
    Per-core L1-L2 BW [B/cy] & $32$  & $64$ \\
    Per-core L2-L3 BW [B/cy] & $32$  & $2\times 16$ \\
    Compiler & Intel 17.0.1 & Intel 18.0.1 \\
    IACA version & 2.1 & 3.0 \\
    \bottomrule
\end{tabular}
\end{table}
\begin{table}
\small\sf\centering
\caption{Useful benchmark values (double precision).
  Execution times for vector operations are
    given \emph{per scalar iteration}. \label{tab_bench}}
    \begin{tabular}{L{2.5cm}ccccc}
        \toprule
 Inverse throughput & \multicolumn{2}{c}{IVB} & \multicolumn{3}{c}{SKX} \\
        \cmidrule(lr){2-3} \cmidrule(lr){4-6}
        for & SSE & AVX & SSE & AVX & AVX512 \\
        \midrule
    vector \texttt{exp()} [cy] & 11.5 & 8.0 & 6.7 & 3.5 & 1.5 \\
        vector \texttt{div}
        \endnote{Values taken from~\cite{fog2017instruction}} [cy] & 7 & 7 & 2 & 2 & 2 \\
        \midrule
    scalar \texttt{exp()} [cy] & \multicolumn{2}{c}{27.8} & \multicolumn{3}{c}{15.1} \\
        \bottomrule
\end{tabular}
\end{table}
The ECM model for the SKX architecture has not been fully developed to date,
but a preliminary formulation based on \eqref{eq_T_ECM} that takes
into account the victim cache architecture of the L3 was
published in~\cite{hager2018applying}.
The heuristics governing cache replacement policies are not disclosed by Intel, but we have found that the following assumptions usually lead to good model predictions:
\begin{itemize}
\item Read traffic from DRAM goes straight to L2.
\item All evicted cache lines from L2, both clean and dirty, are moved to L3.
\item The data path between the L2 and the L3 cache can be assumed to provide
  a bandwidth of 16 B/cy in both directions (i.e., full duplex).
\end{itemize}
The most relevant hardware features for the modeling of both architectures are presented in Tables~\ref{tab_hw} and~\ref{tab_bench}. The Intel IACA tool was used for
estimating in-core execution times of loop kernels. Although IACA supports
both architectures, its support for critical path (CP) prediction was recently dropped.
The IACA outputs for all kernels are available in the reproducibility
appendix.

We illustrate the application of the ECM model to SKX
with the STREAM triad kernel developed by~\cite{McCalpin:1995}:
\begin{equation}\label{eq_stream}
\mathtt{A(:) = B(:) + k* C(:)}~.
\end{equation}
Considering only AVX vectorization as an example, this kernel has the following
properties \emph{per scalar iteration}:
\begin{itemize}
\item AGU-bound inverse throughput prediction of $T_{OL}=0.375$ cy/it 
\item Two loads and one store, so $T_{nOL} = 0.25$ cy/it
\item $V_{L1L2} = 32$ B/it (including write-allocate)
\item $T_{L1L2} = \frac{32\,\mathrm{B/it}}{64\,\mathrm{B/cy}} = 0.5$ cy/it
\item Due to the victim L3 cache, we have to distinguish in-memory and in-L3 datasets.
  \begin{itemize}
  \item L3: $V_{L2L3}^{L3}= 48$ B/it (read + write)
  \item Memory: $V_{L2L3}^{Mem}= 24$ B/it (write-only)
  \end{itemize}
\item The transfer times between L2 and L3 are the same in this particular case because reads and writes to L3 can overlap:
  \begin{itemize}
  \item L3: $T_{L2L3}^{L3}= \max\left(\frac{24\,\mathrm{B/it}}{16\,\mathrm{B/cy}},\frac{24\,\mathrm{B/it}}{16\,\mathrm{B/cy}}\right) = 1.5$ cy/it 
  \item Memory: $T_{L2L3}^{Mem}= \frac{24\,\mathrm{B/it}}{16\,\mathrm{B/cy}} = 1.5$ cy/it (write-only)
  \end{itemize}
\item $V_{L2Mem} = 24$ B/it (read-only traffic)
\item $T_{L2Mem} = \frac{24\,\mathrm{B/it}}{105\,\mathrm{GB/s}}\times2.3\,\mathrm{Gcy/s} = 0.53$ cy/it
\item $V_{L3Mem} = 8$ B/it (write-only traffic)
\item $T_{L3Mem} = \frac{8\,\mathrm{B/it}}{105\,\mathrm{GB/s}}\times2.3\,\mathrm{Gcy/s} = 0.18$ cy/it
\end{itemize}
So the ECM model contributions for the STREAM triad kernel in \eqref{eq_stream} on SKX-AVX would be:
\[
\ecm{T_{OL}}{T_{nOL}}{T_{L1L2}}{T_{L2L3}}{T_{L2Mem}+T_{L3Mem}}{}=
\]\[
\ecm{0.38}{0.25}{0.5}{1.5}{0.71}{\text{cy/it}}\cma
\]
with corresponding predictions according to the non-overlapping machine model of \ecmp{0.38}{0.75}{2.25}{2.96}{\text{cy/it}}. For validation we compared these predictions to benchmark measurements and obtained \ecme{0.39}{0.73}{2.37}{4.3}{\text{cy/it}}, which is in reasonable agreement with the model. The deviation in memory could be fixed by introducing a latency penalty (see~\cite{Hofmann:2017}), but since the memory contribution is rather small for most of the kernels studied here we opted for a simpler model.
In this simple example we have assumed a ``perfect'' machine code with the minimum number of instructions per scalar iteration. For the modeling of more complex kernels we use the actual, unmodified assembly code as generated by the compiler. 

To roughly compare the two architectures a common approach is to use the peak
performance as a metric, measured in single-precision or double-precision \GFS.
The IVB chip supports AVX vectorization and can retire one multiply and one
add instruction per cycle, while the SKX chip supports AVX512 vectorization
and can retire two fused multiply-add instructions per cycle.
This leads to the peak performance numbers shown in
Table~\ref{tab_peak_perf}.
\begin{table}
    \caption{Peak performance for the target
    architectures.\label{tab_peak_perf}}
    \begin{tabular}{lcc}
        \toprule
        Architecture & formula & DP $P_{peak}$ [\GFS] \\ \midrule
        IVB 1 core & $2.2 \times 4 \times 2$ & 17.6\\
        SKX 1 core & $2.3 \times 8 \times 2 \times 2$ & 73.6 \\
        IVB 1 socket & $10~P_{peak}(1~\text{core})$ & 176 \\
        SKX 1 socket & $18~P_{peak}(1~\text{core})$ & 1324.8 \\[0mm]
        \bottomrule
    \end{tabular}
\end{table}
The naive \Rlm\ uses the peak performance of the chip as the
core-bound limit, but often other limitations apply, such as the load
or store throughput between registers and the L1 cache, or pipeline
stalls due to a long critical path or loop-carried dependencies. The
ECM model takes this into account via the
$T_{nOL}$ and $T_{OL}$ runtime contributions, which are
based on an analysis of the actual loop code.

On IVB we used the Intel 17.0.1 compiler with options
\lstinline!-xSSE4.2! and \lstinline!-xAVX! for SSE and AVX code,
respectively.
On SKX we used the Intel 18.0.1 compiler with options
\lstinline!-xSSE4.2!, \lstinline!-xAVX2! and \lstinline!-xCORE-AVX512 -qopt-zmm-usage=high! for SSE, AVX and AVX512 code, respectively.
On both machines we employed 
\lstinline!#pragma ivdep!,~\lstinline!#pragma vector aligned! and 
\lstinline!#pragma omp simd simdlen(N)!
directives where appropriate to ensure vectorization. The
compiler option \lstinline!-qopt-streaming-stores never! was used to
disable the generation of nontemporal stores by the compiler.

To measure relevant performance metrics such as data transfer through the memory hierarchy
we used the \lstinline!likwid-perfctr! tool from the well-established LIKWID framework presented by~\cite{psti,likwidweb}.
We instrumented the code using markers and inserted a barrier before
the execution of each kernel to ensure that measurements would be minimally affected by load imbalance.
On both architectures we employed the CACHES performance group and pinned the OpenMP threads to
the physical cores of a socket. In order to guarantee reproducible benchmark runs, we employed the \lstinline!likwid-setFrequencies! tool to set the clock speed of the cores (and the Uncore in case of SKX) to the base values indicated in Table~\ref{tab_hw}.
In spite of this, we observed the well-known kernel-specific clock frequency throttling on the SKX architecture at all vectorization levels: the average clock frequency never exceeded the limits in Table~\ref{tab_hw} but in some kernels it was observed to be lower, although never less than 2.1\,\GHZ.
In the course of our analysis, we scale our performance predictions by the measured clock frequency whenever required.

\subsection{Simulation of morphologically detailed neurons}
A common approach to modeling morphologically detailed neurons is the so-called
conductance-based (COBA) compartmental model as formalized in the reviews by~\cite{brette2007simulation,bhalla2012multi} and~\cite{gerstner2014neuronal}.  In
this abstraction the arborization of dendrites and axons is represented as a
tree of \emph{sections}, where a section corresponds to an unbranched portion of
the neuron.  Each section is then divided into \emph{compartments} that represent
discretization units for the numerical approximation.  Quantities of interest
such as membrane potential or channel gating variables are typically only
well defined at compartment centers and branching points.

In the compartmental model each compartment is considered analogous to an RC
circuit where the resistance (or rather, the conductance) term can be
nonlinearly dependent on the membrane potential itself.  Due to  their
stability,  implicit methods are a common choice for
time integration of the differential equations arising from this representation,
thus the solution of a linear system of equations is required at each
time step.  In the presence of branching points, this leads to a
quasi-tridiagonal system that can be solved in linear time using the algorithm
proposed in~\cite{hines1984efficient}.

\begin{figure*}
\centering
    \includegraphics[width=0.99\textwidth]{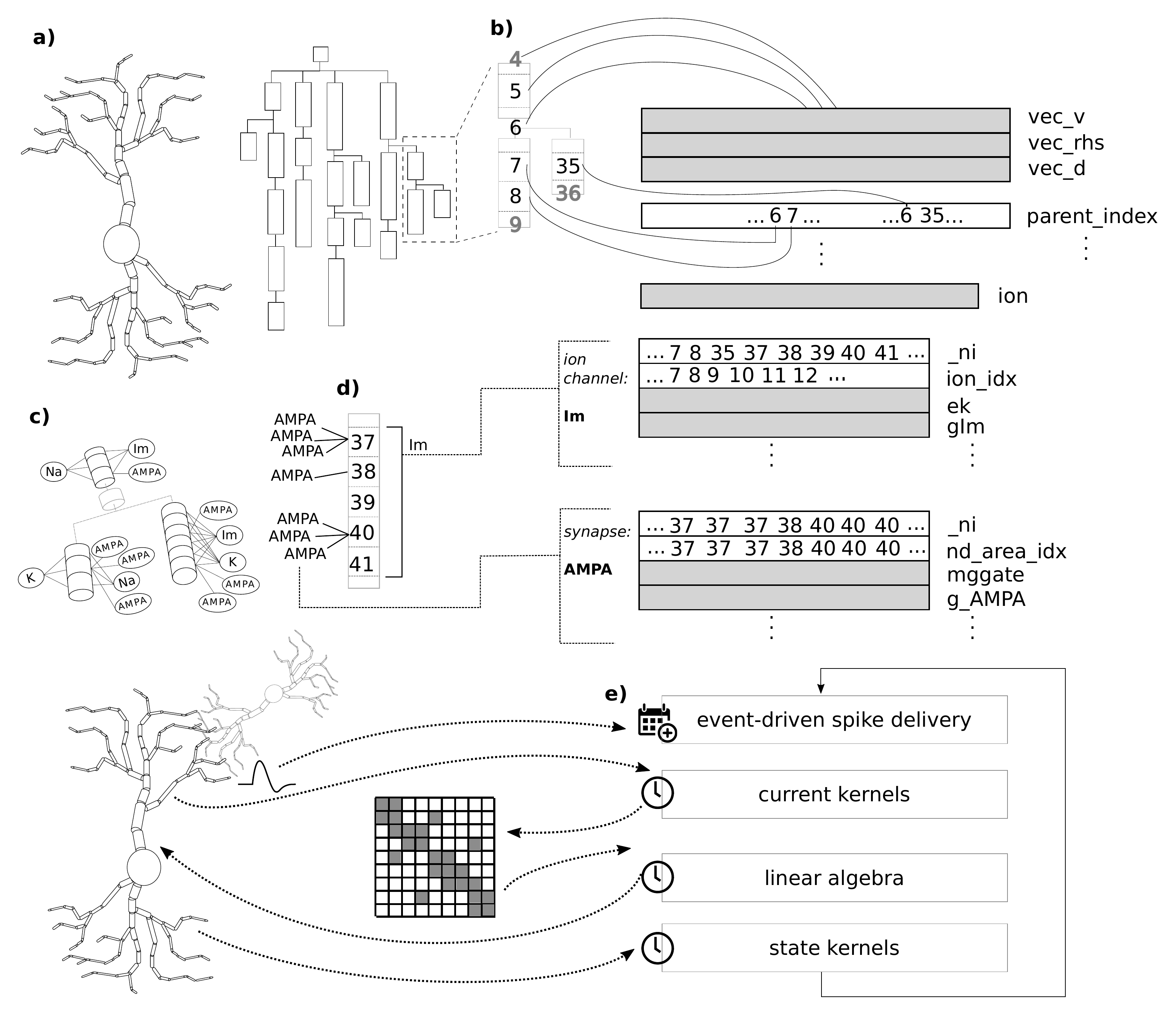}
    \caption{Neuron representation and data layout.
    \textbf{a):} Neurons are represented as a tree of unbranched sections, where each section can be further split into compartments for numerical discretization.
    \textbf{b):} Each compartment is numbered according to the schema in~\cite{hines1984efficient}, and the tree structure is represented in memory by an array of \lstinline!parent_index!.
    Additional arrays are used to represent the neuron's state (e.g. \lstinline!vec_v! holds the membrane potential of each compartment), and three arrays are used for a sparse representation of the time integration matrix.
    Arrays of double precision values are colored in grey, while arrays of integer indices are white and contain some elements to give an idea of their structure.
    \textbf{c):} Additionally, every compartment can be endowed with zero or more ion channels or synapses, which require additional arrays to be represented.
    Branching points (in grey) are treated as any other compartments for the purposes of linear algebra, but cannot have any instances of ion channels or synapses.
    \textbf{d):} Ion channels (e.g., \lstinline!Im!) either have a single instance in all the compartments of a section, or do not have any instances at all in that section.
    Synapses (e.g. \lstinline!AMPA!) can have multiple instances per compartment and do not need to be represented in all the compartments belonging to the same section.
    \textbf{e):} The application's workflow, excluding bookkeeping and parallel communication.
    First, the spike delivery kernel is called only for all the events that have been generated by other neurons and that have an effect on synapses of this neuron;
    then, at every timestep, the current, linear algebra and state kernels are executed.
    Current kernels read information from the state of the neuron and update the linear system's matrix.
    Linear algebra solves the linear system using a custom method and updates the state of the membrane potential.
    State kernels read the membrane potential and update the state of all the ion channels and synapses.
    \label{fig_data_layout}}
\end{figure*}

In the COBA model, the membrane conductance is determined by aggregating several
contributions from ion channels, which are special cross-membrane proteins that
allow an ionic current to flow into or out of the
cell.  Thus in the COBA compartmental model, when using an implicit time
integrator, three algorithmic phases are required to advance a neuron in time:
first one must compute the contributions to the linear system (the ion channel
and synapses \emph{currents}); then one must solve the linear system; finally, one must
update the \emph{states} of individual ion channel and synapse instances based on the
recently-computed compartment potentials (see Figure~\ref{fig_data_layout}).

Neurons also have the ability to communicate with other neurons using synapses:
points of contact between different neurons that are triggered when an action potential is elicited in the presynaptic cell and, at the onset of this event, determine a change in the membrane potential of the postynaptic cell.  Therefore the simulation
algorithm is composed of two parts: a clock-driven portion that advances the
state of a neuron from a timestep to the next; and an event-driven part that is
only executed when a synaptic event is received.
Figure~\ref{fig_data_layout} presents a summary of the main algorithm phases and data
layout.

The compartmental modeling of neurons using COBA formalism
is implemented in the widely-adopted software for neuroscientific simulations
NEURON.  The NEURON software is a long-lasting
project that includes an interpreter for a custom scripting language (HOC),
a domain specific language tool to expand the models of ion channels and synapses, a
GUI and a domain specific language (NMODL) to expand the repertoire of available models.
To reduce the complexity and concentrate on
the fundamental computational properties of the simulation kernels, in this work
we utilize instead CoreNEURON, a lean version of NEURON's
simulation engine based on the work by~\cite{kumbhar2016leveraging}.
CoreNEURON implements several optimizations over NEURON,
including improved memory requirements and vectorization, at the cost of 
functionality.  In particular, NEURON is usually still required to define a
model and a simulation setup before CoreNEURON executes the simulation.
The NEURON/CoreNEURON software allows neuroscientists to specify custom ion channel and synapse models using the domain specific language NMODL introduced in~\cite{hines2000expanding}, which is then automatically translated into C code and compiled in a dynamic library.

The CoreNEURON data layout is shown in Figure~\ref{fig_data_layout}.
First the neuron is modeled logically as a tree of unbranched sections, whose topology is represented by a vector of parent indices.
Other relevant quantities such as the membrane potential and the tridiagonal sparse matrix are are represented by double precision arrays with length equal to the number of compartments.
More details about the matrix representation are given in Section~\ref{sec_linalg}.
Additionally, ion channel-specific and synapse-specific quantities are held in separate data structures consisting of arrays of double precision values in Structure-of-Arrays layout (SoA), indices to the corresponding compartments and, if needed, pointers to other internal data structures.
\subsection{Preliminary performance observations and motivation\label{sec_prelim}}
\begin{figure*}
    \centering
    \includegraphics[width=0.95\textwidth]{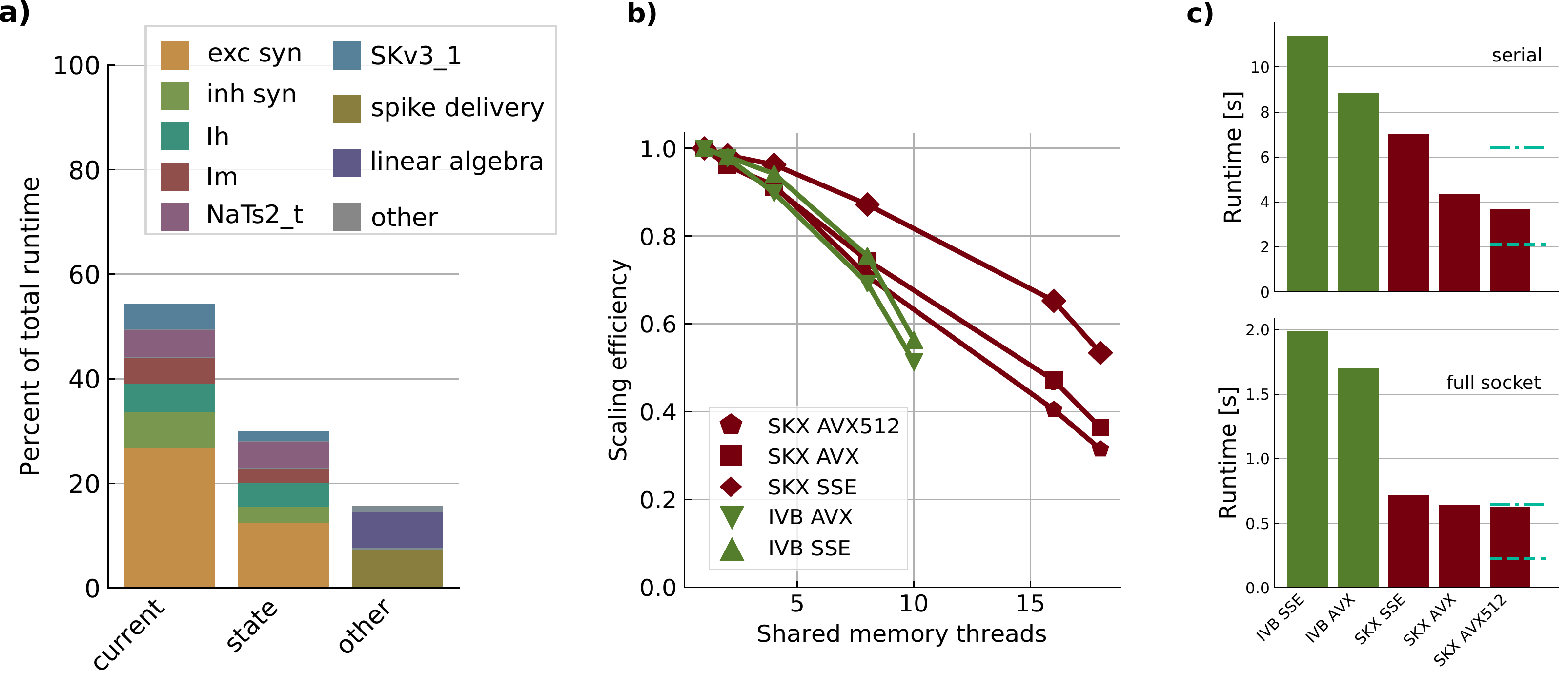}
    \caption{Measured performance and observations from benchmark.
        \textbf{a):} breakdown of the distribution of relative importance of the different kernels in the simulation of a full neuron for the SKX-AVX512 architecture using a full socket.
    Overhead from the rest of the execution is not shown.
    Linear algebra and spike delivery combined hardly exceed 10\%.\label{fig_breakdown}
    \textbf{b):} median strong scaling efficiency over 10 runs.
    Measurements exhibited little variability across different runs, so quantile errorbars are not visible.
    Parallel efficiency degrades quickly, especially on SKX, and vectorization strengthens this negative effect.
    \label{fig_strong_efficiency}
    \textbf{c):} total runtime to simulate one neuron for one second in the serial case (top) and using the full socket (bottom).
    Overhead from the non-computational kernels is not considered.
    The dashed blue line represents the expected runtime if scaled perfectly with the architecture's theoretical peak performance from IVB-AVX to SKX-AVX512, while the dash-dotted blue line marks the expected runtime if scaled perfectly with the ratio of measured memory bandwidths.
    We do not observe the ideal speedup, and in this paper we employ performance modeling to explain the underlying reasons.
    \label{fig_meas-runtime-vect}
    }
\end{figure*}

Given that the simulation algorithm is composed of many phases with different characteristics, the first step in performance analysis is a search for the time-consuming hot spots. 
In the shared-memory parallel execution, \emph{current} and \emph{state} kernels are usually dominant, representing roughly 80\% of the total execution time, while the event-driven \emph{spike delivery} and \emph{linear algebra} kernels account for 10--20\% (see Figure~\ref{fig_breakdown}a).
In the serial execution we observe that the relative importance of spike delivery increases slightly, however, the state and current kernels still dominate.
This serial performance profile was also observed in~\cite{kumbhar2016leveraging} and is a peculiar feature of compartmental COBA models, whereas current-based point neuron models are typically dominated in serial execution by event-driven spike delivery and event bookkeeping, as shown in the work by~\cite{peyser2015nest}.
Unfortunately, these results are tightly linked to the benchmark setup, and it is unknown whether this is a general property of COBA models or not.

We have chosen two Intel architectures that were introduced about five years apart in order to be able to identify the speedup from architectural improvements.
Judging by the peak performance numbers in Table~\ref{tab_peak_perf} one would expect a per-socket speedup of about 7.5$\times$. On the other hand, comparing memory bandwidth (see Table~\ref{tab_hw}), which is the other lowest-order bottleneck of code execution, a factor of 2.6$\times$ could be estimated. As shown in Figure~\ref{fig_breakdown}c, we observe a factor of roughly 3$\times$ between the best IVB and SKX versions. Although it is satisfying that the measurement lies between the two estimates, detailed performance modeling is required to explain the \emph{actual} value. 

One of the main in-core features of modern architectures is the possibility to expose data parallelism using vectorized (SIMD) instructions on wide registers.
We investigated the benefits of vectorization at different levels of thread-parallelism.
In the serial execution (see Figure~\ref{fig_breakdown}c) we found that the Skylake architecture had in general better performance than Ivy Bridge, and that using wider registers improved the performance, even though the acceleration was not ideal (i.e., not in line with the larger register width).
At full socket we found that the difference between architectures was exacerbated, while we saw only minor improvements from vectorization (see Figure~\ref{fig_breakdown}c).
We also investigated the strong-scaling efficiency of the simulation code on different architectures (Figure~\ref{fig_breakdown}b) and found that, as expected, the efficiency decreases as the level of parallelism grows.
This indicates a tradeoff in terms of chip and software design: further analysis is required to understand whether it is worth investing in SIMD or shared-memory parallelism, optimize for instruction level parallelism, out of order execution or a combination of all of these.

We exploit performance modeling techniques in order to gain insight into the interaction between the CoreNEURON simulation code and modern hardware architectures.
This will allow us to answer the open performance questions above as well as to generalize to different architectures for future co-design efforts.
\section{Performance Modeling of Detailed Simulations of Neurons}\label{sec:modeling}
\subsection{Ion channels current kernels}\label{sec_ion_current}
Ion channel \emph{current kernels} are used in the simulation algorithm to update the matrix representing the voltage equation by computing contributions from the ionic current of different chemical species.
We consider in this work four ion channel types that are among the most representative: \lstinline!Ih, Im, NaTs2_t, SKv3_1!.
In Listing~\ref{lst_Im_current} we show the code for the \lstinline!Im current! kernel as an example; other kernels share similar memory access patterns and arithmetic operations with only minor changes.
\begin{lstlisting}[language=C, tabsize=2, basicstyle=\footnotesize\ttfamily,float,
caption={\lstinline{Im current} kernel},label={lst_Im_current}]
for(int i=0; i<cntml; ++i) {
  int nd_idx = _ni[i];
  double v = vec_v[nd_idx];
  ek[i] = ion_data_ek[ion_idx[i]];
  gIm[i] = gImbar[i] * m[i] ;
  ik[i] = gIm[i] * ( v - ek[i] ) ;
  ion_data_ik[ion_idx[i]] += ik[i] ;
  vec_rhs[nd_idx[i]] -= ik[i];
  vec_d[nd_idx[i]] += gIm[i];
}
\end{lstlisting}

\emph{Current kernels} are typically characterized by two main features: low arithmetic intensity and scattered loads/stores.
The latter can present a modeling problem, but in practice we can obtain good accuracy using a few heuristics based on domain-specific knowledge.
In particular, as a first approximation one can treat the indices in \lstinline!_ni! and \lstinline!ion_idx! as perfectly contiguous (see Figure~\ref{fig_data_layout} as a justification).
In total, the kernel reads from four double and two integer arrays, and writes to six double arrays, leading to 136\,\byte\ of overall data traffic per scalar iteration through the complete memory hierarchy (this includes write-allocate transfers on store misses).

Combining the data volume estimates with in-core predictions from IACA (using the full throughput assumption) we can generate the ECM model predictions in cycles per scalar iteration as shown in Table~\ref{tab_Im_curr}.
The compiler is able to employ scatter/gather instructions for this kernel on SKX (these are not supported on IVB).
As expected, the model predicts that the performance of this strongly data-bound kernel will degrade as the data resides farther from the core. Vectorization is not beneficial at all except for AVX512 with data in L1, which can be attributed to the required scalar load instructions when gather/scatter instructions are missing. 
To validate the predictions we designed a serial benchmark that allowed fine-grained control over the dataset size by removing all ion channels and synapses except \lstinline!Im! from our dataset, but still executing the complete application loop.
The resulting dataset size was roughly 50\,\KB\ and 200\,\KB\ for the L2 benchmarks and 6\,\MB\ and 7\,\MB\ for the L3 benchmarks on the IVB and SKX architectures, respectively.
Due to overheads, it was impossible to construct a benchmark for the L1 cache on either machine.

On IVB the measurements remained within 10\% of the predictions for all levels of the memory hierarchy, while on SKX the ECM predictions were a little more off, especially for data in the cache. This might be caused by our simplifying model assumptions about the data transfers between L2 and L3, for which no official documentation exists. 
Still the ECM model gave accurate predictions in almost all of our benchmarks and provided insight into the computational properties of this kernel.

We conclude that the \lstinline!Im current! kernel, and all \emph{current kernels} in general, are data-bound and limited solely by data transfer capabilities of the system across the memory hierarchy.
Even for an in-memory dataset, wider data paths between the caches would thus benefit the performance of the kernel. The clock frequency will have a significant but weaker than linear impact on the performance because memory transfer rates are only weakly dependent on it (especially on the more modern architectures like SKX).
The analysis also predicts strong memory bandwidth saturation with a few (4--5) cores, so the memory bandwidth starts to play a decisive role once bandwidth saturation is achieved.
\begin{table*}
    \centering
    \caption{ECM model and serial measurements per scalar iteration [cy/it] for the \lstinline{Im current} kernel. \label{tab_Im_curr}}
\begin{tabular}{lccc}
    \toprule
    & contributions & predictions & measurements \\[0.5mm]
    \midrule
    IVB SSE & \ecm{7.8}{5.5}{4.2}{4.2}{7.5}{}
            & \ecmp{7.8 }{9.7 }{13.9 }{21.4}{}
            & \ecme{\text{n/a} }{10.8 \pm 0.1 }{13.5 \pm 0.0 }{23.8 \pm 0.1}{}\\[0.5mm]
    IVB AVX & \ecm{7.8}{5.6}{4.2}{4.2}{7.5}{}
            & \ecmp{7.8 }{9.8 }{14.0 }{21.5}{}
            & \ecme{\text{n/a} }{10.2 \pm 0.0 }{13.1 \pm 0.0 }{23.8 \pm 0.2}{}\\[0.5mm]
	SKX SSE & \ecm{7.8}{5.5}{2.1}{5.5}{3.0}{}
			& \ecmp{7.8}{7.8}{13.1}{16.1}{}
            & \ecme{\text{n/a} }{9.1 \pm 0.1 }{11.0 \pm 1.0 }{15.3 \pm 1.0}{}\\[0.5mm]
	SKX AVX & \ecm{7.3}{4.8}{2.1}{5.5}{3.0}{}
			& \ecmp{7.3}{7.3}{12.4}{15.4}{}
            & \ecme{\text{n/a} }{8.7 \pm 0.1 }{11.4 \pm 0.0 }{15.0 \pm 1.2 }{}\\[0.5mm]
	SKX AVX512 & \ecm{5.3}{3.0}{2.1}{5.5}{3.0}{}
			& \ecmp{5.3}{5.3}{10.6}{13.6}{}
            & \ecme{\text{n/a} }{7.6 \pm 0.0 }{10.6 \pm 0.8 }{15.6 \pm 1.5}{}\\[0.5mm]
    \bottomrule
\end{tabular}
\end{table*}
\subsection{Synaptic current kernels}\label{sec_syn_cur}
Synapses are arguably the pivotal component of neuron simulations.
Synaptic current kernels are particularly important for performance as shown in Figure~\ref{fig_breakdown}, and pose a modeling challenge because of their complex chain of intra-loop dependencies, memory accesses and presence of transcendental functions.
There are two types of synapses in this dataset: excitatory AMPA/NMDA synapses and inhibitory GABAAB synapses.
As an example, the source code for the excitatory AMPA/NMDA synapse current is shown in Listing~\ref{lst_syn_cur}.
\begin{lstlisting}[language=C, tabsize=2, basicstyle=\footnotesize\ttfamily,float,
caption={Excitatory synapse current kernel}, label={lst_syn_cur}]
for(int i=0; i<cntml; ++i) {
  double v = vec_v[_ni[i]];
  mggate[i] = 1.0 + exp (-0.062*v)*(mg[i]/3.57);
  mggate[i] = 1.0/mggate[i];
  g_AMPA[i] = gmax * ( B_AMPA[i] - A_AMPA[i] ) ;
  g_NMDA[i] = gmax * ( B_NMDA[i] - A_NMDA[i] ) ;
  g_NMDA[i] *= mggate[i] ;
  g[i] = g_AMPA[i] + g_NMDA[i] ;
  i_AMPA[i] = g_AMPA[i] * ( v - e[i] ) ;
  i_NMDA[i] = g_NMDA[i] * ( v - e[i] ) ;
  i_tot[i] = i_AMPA[i] + i_NMDA[i] ;
  double rhs = i_tot[i];
  double _mfact =  1.e2/(_nd_area[nd_area_idx[i]]); 
  double loc_g =  g_AMPA[i] + g_NMDA[i] ;
  loc_g *= _mfact;
  rhs *= _mfact;
  vec_shadow_rhs[i] = rhs;
  vec_shadow_d[i] = loc_g;
}
\end{lstlisting}
The expensive exponentials and divides in this code are balanced by large data requirements. The kernel reads one element each from eight double and two integer arrays, and writes one element each to nine double arrays, which would amount to a traffic of 216\,\byte\ per iteration.
However, as shown in Figure~\ref{fig_data_layout}, the typical structure of the \lstinline!_ni! and \lstinline!nd_area_idx! arrays is different from that of the indexing arrays in ion channel kernels.
In particular, as a direct consequence of multiple synapse instances being able to coexist within the same compartment, the \lstinline!_ni! and \lstinline!nd_area_idx! arrays often exhibit sequences of repeated elements.
This means that subsequent iterations of the kernel can exploit some temporal locality in accessing the \lstinline!vec_v! and \lstinline!_nd_area! arrays.
To account for this we reduce the expected traffic from these arrays by a weighting factor  equal to the average length of a sequence of repeated elements in \lstinline!_ni! and \lstinline!nd_area_idx!, which is about 3 in our case.
Thus the updated data traffic estimate is 205\,\byte\ through the complete memory hierarchy.
To compute $T_{OL}$ the inverse throughput of the vectorized exponential operation from Table~\ref{tab_bench} must be added to the kernel runtime reported by IACA, and $T_{nOL}$ is derived from the retired load instructions as usual. 
We then obtain the ECM predictions per scalar iteration in Table~\ref{tab_syn_curr}.

The analysis reveals a complex situation.  Both code versions on IVB
and the SSE code on SKX are predicted to be core bound as long as the data fits into any
cache. The AVX and AVX512 code on SKX, however, become data bound already
in the L3 cache.

Again we used a benchmark dataset containing only synapses to validate the model, with a  size of roughly 80\,\KB\ and 500\,\KB\ for the L2 benchmarks and 1.5\,\MB\ and 11\,\MB\ for the L3 benchmark on the IVB and SKX architectures, respectively.
On both CPUs the model predictions are optimistic compared to measurements by a 10--50\% margin.
Interestingly, within each architecture the model becomes more accurate as the SIMD width increases.
Even though the predictions are not all within a small accuracy window, the model still allows us to correctly categorize the relevant bottlenecks, and is especially effective in capturing the fact that on SKX with AVX512 the kernel is rather strongly data bound.
Given the complex inter-dependencies between operations in the kernel, we speculate that a critical path might be invalidating the full-throughput assumption of the ECM model, although this would not be sufficient to explain why the DRAM measurements are larger than the L2 and L3 measurements.

As a result from the analysis we conclude that, for an in-memory dataset, the performance of the serial excitatory synapse current kernel would improve significantly only if in-core execution and data transfers were enhanced at the same time. Still the model predicts bandwidth saturation for all code variants, once run in parallel, at 4--6 cores. 
\begin{table*}
    \centering
    \caption{ECM model and serial measurements per scalar iteration [cy] for the excitatory  synapse current kernel.\label{tab_syn_curr}}
    \begin{tabular}{lccc}
        \toprule
        & contributions & predictions & measurements \\[0.5mm]
        \midrule
	IVB SSE & \ecm{32.5}{9.8}{6.4}{6.4}{11.3}{}
            & \ecmp{32.5 }{32.5 }{32.5 }{33.9 }{}
            & \ecme{\text{n/a} }{39.6 \pm 0.2 }{39.4 \pm 0.0 }{44.0 \pm 0.2 }{}\\[0.5mm]
	IVB AVX & \ecm{29.0}{7.8}{6.4}{6.4}{11.3}{}
            & \ecmp{29.0 }{29.0 }{29.0 }{31.9 }{}
            & \ecme{\text{n/a} }{32.9 \pm 0.1 }{33.0 \pm 0.1 }{36.1 \pm 1.6 }{}\\[0.5mm]
	SKX SSE & \ecm{21.6}{9.9}{3.2}{8.3}{4.5}{}
			& \ecmp{21.6}{21.6}{21.6}{25.9}{}
            & \ecme{\text{n/a} }{31.3 \pm 0.1 }{31.4  \pm 0.1 }{32.2 \pm 0.0 }{}\\[0.5mm]
	SKX AVX & \ecm{13.5}{7.0}{3.2}{8.3}{4.5}{}
			& \ecmp{13.5}{13.5}{18.5}{23.0}{}
            & \ecme{\text{n/a} }{16.9 \pm 0.1 }{17.0  \pm 0.5 }{23.9 \pm 3.5 }{}\\[0.5mm]
	SKX AVX512 & \ecm{7.2}{3.5}{3.2}{8.3}{4.5}{}
			& \ecmp{7.2}{7.2}{15.0}{19.5}{}
            & \ecme{\text{n/a} }{10.9 \pm 0.1 }{13.5  \pm 0.8 }{25.1 \pm 1.9 }{}\\[0.5mm]
        \bottomrule
    \end{tabular}
\end{table*}
\subsection{Ion channels state kernels}\label{sec_ion_state}
During the execution of a \emph{state kernel}, the internal state variables of an instance of an ion channel or a synapse are integrated in time and advanced to the next timestep.
Figure~\ref{fig_breakdown}a shows that state kernels represent a significant portion of the overall runtime, although their relative importance is largest in the single-thread execution and decreases with shared-memory parallelism.

State kernels are characterized by a very large overlapping contribution $T_{OL}$ due to exponential functions and division operations, combined with low data requirements. This gives reason to expect a clearly core-bound situation.
As an example, we show the code for the \lstinline!Ih! state kernel in Listing~\ref{lst_Ih_state}.
\begin{lstlisting}[language=C, tabsize=2, basicstyle=\footnotesize\ttfamily,float,
caption={Ih state kernel}, label={lst_Ih_state}]
for(int i=0; i<cntml; ++i) {
  double v = vec_v[_ni[i]];
  mAlpha[i] = 6.43e-3*(v + 154.9);
  mAlpha[i] /= exp((v + 154.9)/11.9)-1.;
  mBeta[i] = 0.193*exp(v/33.1);
  mInf[i] = mAlpha[i]/(mAlpha[i]+mBeta[i]);
  mTau[i] = 1./(mAlpha[i]+mBeta[i]) ;
  double incr = (1-exp(-dt/mTau[i]));
  incr *= (mInf[i]/mTau[i])/(1./mTau[i]) - m[i];
  m[i] += incr;
}
\end{lstlisting}
In analogy with the previous ion channel example, we treat the indices in \lstinline!_ni! as  contiguous.
Therefore this kernel requires reading one element each from one double and one integer array, and writing one element each to three double arrays, amounting to a traffic of 60\,\byte\ per iteration.
On the other hand, the kernel needs three exponential function evaluations and eight divides, of which
some might be eliminated by compiler optimizations (common subexpression elimination and substitution of multiple divides by the same denominator for a reciprocal and several multiplications).

Again combining the IACA prediction with measured throughput data for \lstinline!exp()! (see Table~\ref{tab_bench}) and the data delay we arrive at the ECM predictions per scalar iteration in Table~\ref{tab_Ih_state}.
State kernels can be considered as the polar opposite of current kernels in terms of their computational profile, and the model predicts that their performance will be independent of the location of the working set in the memory hierarchy. This also leads to the expectation that vectorization should yield massive improvements, but the ECM model says otherwise.
According to the performance model these kernels are dominated by the throughput of the \lstinline!exp! function and the eight divides, by comparable amounts; for instance, the SKX-AVX version spends 16\,\cycles\ in divides and another 10.4\,\cycles\ in \lstinline!exp()!. No optimizations concerning the divides are done by the compiler, although the number of divides may be reduced to three by the methods mentioned above.

Both architectures show only moderate speedup from SSE to AVX (13\% on IVB and 37\% on SKX, respectively)\@. On IVB this can be partly attributed to the mere 44\% speedup for the \lstinline!exp()! function (see Table~\ref{tab_bench}), but the main cause on both CPUs is the constant throughput per divide operation, independent of the SIMD width. This is a well-known design tradeoff in Intel architectures: putting a large number of low-throughput units on a core does not pay off on a general-purpose CPU.

AVX512, on the other hand, exhibits a large speedup that cannot be explained by the above analysis. Inspection of the assembly code reveals that the compiler did not generate any divide instructions at all. Instead, it uses \lstinline!vrcp14d! instructions together with Newton-Raphson steps for better throughput on SKX (see \cite{ia32opt:2018})\@.

We validated our predictions with dataset sizes of 124\,\KB\ and 500\,\KB\ for the L2 benchmarks on IVB and SKX respectively, and a dataset size of 5\,\MB\ for the L3 benchmarks on both architectures.
Except for the AVX kernels, for which the accuracy is more than satisfying, the predictions are optimistic by between 15\% and 35\%. It must be stressed that when a loop is strongly core bound and has a long critical path, the automatic out-of-order execution engine in the hardware may have a hard time overlapping successive loop iterations. Since the ECM model has no concept of this issue, predictions may be qualitative.

Despite all inaccuracies, the conclusion from the analysis is clear: Faster exponential functions,
wider SIMD execution for divide instructions and a higher clock frequency would immediately (and proportionally) boost the performance of the serial \lstinline!Ih! state kernel. Memory bandwidth saturation is not expected on IVB, but on SKX the AVX and AVX512 versions will be able to hit the memory bandwidth limit, albeit at a larger number of cores than with the more data-bound kernels. Hence, boosting parallel performance is achieved by different means on the two chips.
\begin{table*}
    \centering
    \caption{ECM model and serial measurements per scalar iteration [cy] for Ih state kernel. \label{tab_Ih_state}}
    \begin{tabular}{lccc}
        \toprule
        & contributions & predictions & measurements \\[0.5mm]
        \midrule
	IVB SSE & \ecm{90.5}{4.5}{2.9}{2.9}{5.1}{}
            & \ecmp{90.5 }{90.5 }{90.5 }{90.5 }{}
            & \ecme{\text{n/a} }{106.7 \pm 0.1 }{106.5 \pm 0.0 }{107.0 \pm 0.0 }{}\\[0.5mm]
	IVB AVX & \ecm{80.0}{4.5}{2.9}{2.9}{5.1}{}
            & \ecmp{80.0 }{80.0 }{80.0 }{80.0 }{}
            & \ecme{\text{n/a} }{80.1 \pm 0.1 }{80.0 \pm 0.1 }{81.9 \pm 0.1 }{}\\[0.5mm]
	SKX SSE & \ecm{36.1}{6.0}{1.4}{3.2}{2.0}{}
			& \ecmp{36.1}{36.1}{36.1}{36.1}{}
            & \ecme{\text{n/a} }{53.4 \pm 0.2 }{53.4 \pm 0.1 }{52.3 \pm 0.0 }{}\\[0.5mm]
	SKX AVX & \ecm{26.4}{3.4}{1.4}{3.2}{2.0}{}
			& \ecmp{26.4}{26.4}{26.4}{26.4}{}
            & \ecme{\text{n/a} }{29.9 \pm 0.1 }{29.9 \pm 0.1 }{28.8 \pm 0.0 }{}\\[0.5mm]
	SKX AVX512 & \ecm{12.1}{1.9}{1.4}{3.2}{2.0}{}
			& \ecmp{12.1}{12.1}{12.1}{12.1}{}
            & \ecme{\text{n/a} }{18.6 \pm 0.1 }{18.3 \pm 0.1 }{19.0 \pm 0.1 }{}\\[0.5mm]
        \bottomrule
    \end{tabular}
\end{table*}
\subsection{Synaptic state kernels}\label{sec_syn_state}
Synapse state kernels have computational properties similar to ion channel state kernels, i.e., a dominating in-core overlapping contribution due to exponentials and divides, coupled with low data requirements.
As an example, we show the code for the excitatory AMPA/NMDA synapse in Listing~\ref{lst_syn_state}.
This kernel reads one element each from four double arrays and updates one element each from four other double arrays, thus totaling 96\,\byte\ of data volume per iteration.
\begin{lstlisting}[language=C, tabsize=2, basicstyle=\footnotesize\ttfamily,float,
caption={Excitatory synapse state kernel}, label={lst_syn_state}]
for(int i=0; i<cntml; ++i) {
  double inc_AA=(1.-exp(dt*(-1./tau_r_AMPA[i])));
  inc_AA *= (-A_AMPA[i]);
  double inc_BA=(1.-exp(dt*(-1./tau_d_AMPA[i])));
  inc_BA *= (-B_AMPA[i]);
  double inc_AN=(1.-exp(dt*(-1./tau_r_NMDA[i])));
  inc_AN *= (-A_NMDA[i]);
  double inc_BN=(1.-exp(dt*(-1./tau_d_NMDA[i])));
  inc_BN *= (-B_NMDA[i]);
  A_AMPA[i]+=inc_AA;
  B_AMPA[i]+=inc_BA;
  A_NMDA[i]+=inc_AN;
  B_NMDA[i]+=inc_BN;
}
\end{lstlisting}
The ECM predictions per scalar iteration are listed in Table~\ref{tab_syn_state}.
An important observation to be made here is that using the \lstinline!AVX2! instruction set was crucial to obtaining good performance on Skylake-X.
Indeed the \lstinline!exp! function invoked by the \lstinline!AVX! instruction set has a much worse throughput (despite having the same vector width) and thus would significantly degrade the performance of this kernel.
As expected, all other observations and conclusions are the same as for the ion channel state kernels in the previous section. All predictions are optimistic by 20--30\%.
\begin{table*}
    \centering
    \caption{ECM model and serial measurements per scalar iteration [cy] for the excitatory synapse state kernel.\label{tab_syn_state}}
    \begin{tabular}{lccc}
        \toprule
        & contributions & predictions & measurements \\[0.5mm]
        \midrule
	IVB SSE & \ecm{75.0}{5.0}{3.0}{3.0}{5.3}{}
            & \ecmp{75.0 }{75.0 }{75.0 }{75.0 }{}
            & \ecme{\text{n/a} }{93.0 \pm 0.1 }{92.7 \pm 0.0 }{94.3 \pm 0.0 }{}\\[0.5mm]
	IVB AVX & \ecm{60.0}{3.9}{3.0}{3.0}{5.3}{}
            & \ecmp{60.0 }{60.0 }{60.0 }{60.0 }{}
            & \ecme{\text{n/a}}{75.0 \pm 0.0}{74.9 \pm 0.0}{75.0 \pm 0.4}{}\\[0.5mm]
	SKX SSE & \ecm{34.8}{6.5}{1.5}{4.0}{2.1}{}
            & \ecmp{34.8}{34.8}{34.8}{34.8}{}
            & \ecme{\text{n/a} }{45.7 \pm 0.0 }{45.7  \pm 0.0 }{44.9 \pm 0.0 }{}\\[0.5mm]
	SKX AVX & \ecm{22.0}{3.8}{1.5}{4.0}{2.1}{}
            & \ecmp{22.0}{22.0}{22.0}{22.0}{}
            & \ecme{\text{n/a}}{25.5 \pm 0.1}{25.5 \pm 0.1}{25.7 \pm 0.2}{} \\[0.5mm]
	SKX AVX512 & \ecm{9.7}{1.7}{1.5}{4.0}{2.1}{}
            & \ecmp{9.7}{9.7}{9.7}{9.7}{}
            & \ecme{\text{n/a} }{13.1 \pm 0.1 }{13.4 \pm 0.2 }{13.7 \pm 0.2 }{}\\[0.5mm]
        \bottomrule
    \end{tabular}
\end{table*}
\subsection{Validation for all state and current kernels}
To assess the validity of our performance and conclusions about bandwidth saturation on a real-world use case we designed a representative dataset based on the \lstinline!L5_TTPC1_cADpyr232_1! neuron, which can be downloaded from the Blue Brain NMC portal introduced in~\cite{ramaswamy2015neocortical}.
Since L5 pyramidal cells are among the cell types with the largest computational load in the reconstruction of the rat neocortex by~\cite{markram2015reconstruction}, this constitutes a highly representative subset of a full cortical column reconstruction.
Commonly studied network arrangements are composed of a large number of neurons to be able to capture macroscopic effects, and even in the case of distributed simulations this usually amounts to a few hundred or even thousands of neurons per node.
Given that the average detailed neuron among those in the Blue Brain NMC portal requires roughly 2\,\MB\ of data, this means that one can usually assume that data must be fetched from main memory each time.
We used a sufficiently large dataset consisting of 500 copies of the neuron mentioned above (for a total of 850\,\MB) as a building block for our benchmarks, eventually duplicating it according to the type of scaling scenario under analysis to avoid load imbalance issues.

\begin{figure*}
    \centering
    \includegraphics[width=0.195\textwidth]{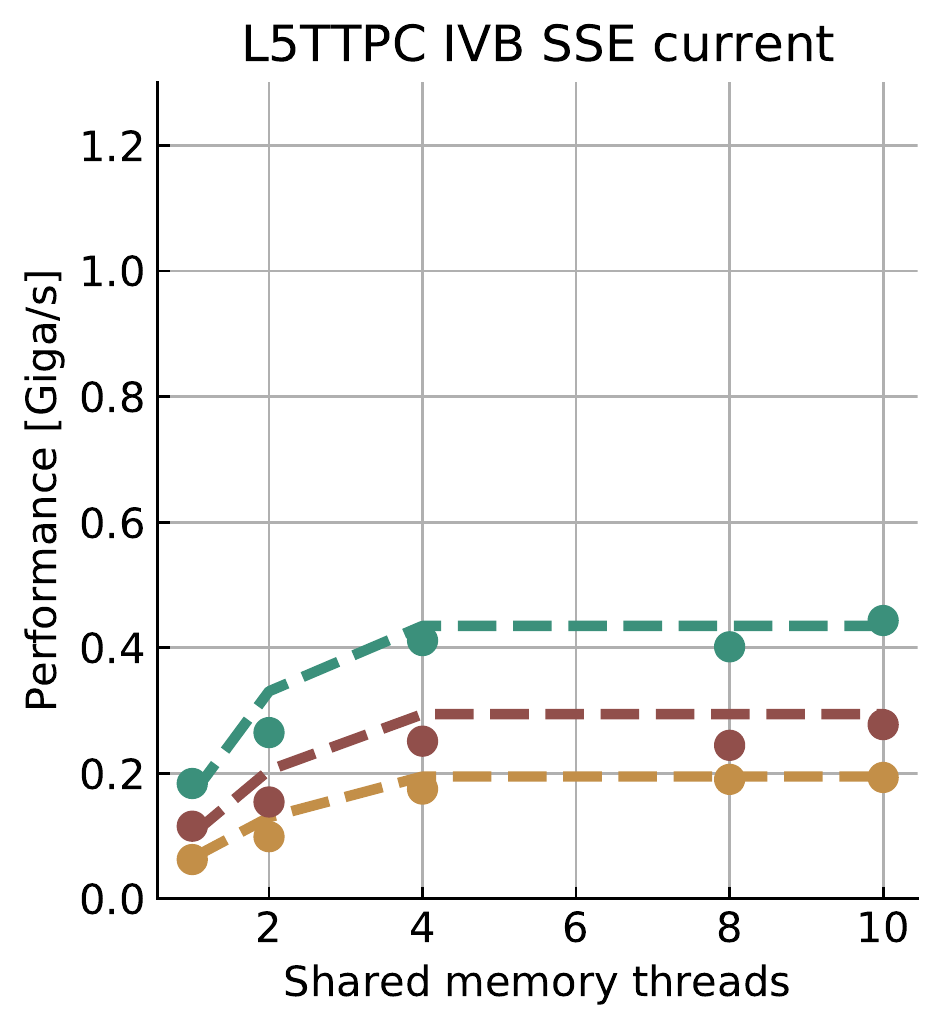}~
    \includegraphics[width=0.195\textwidth]{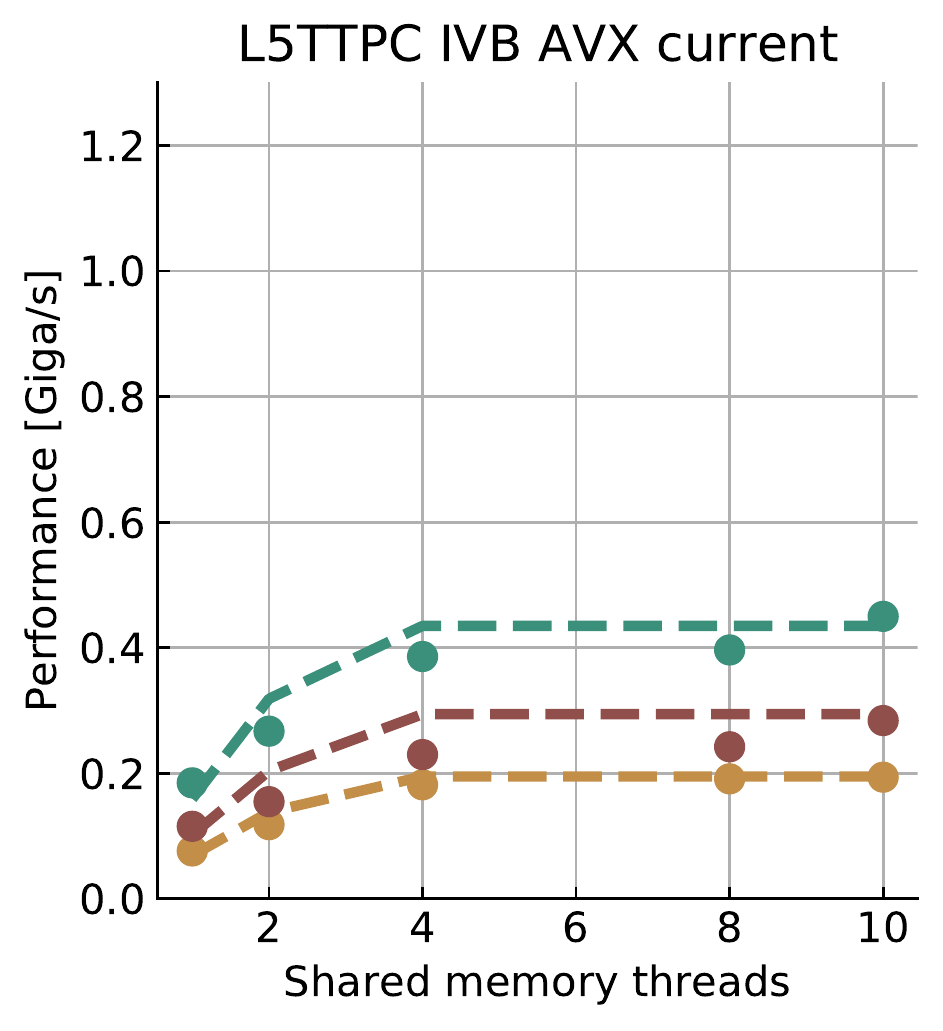}~
    \includegraphics[width=0.195\textwidth]{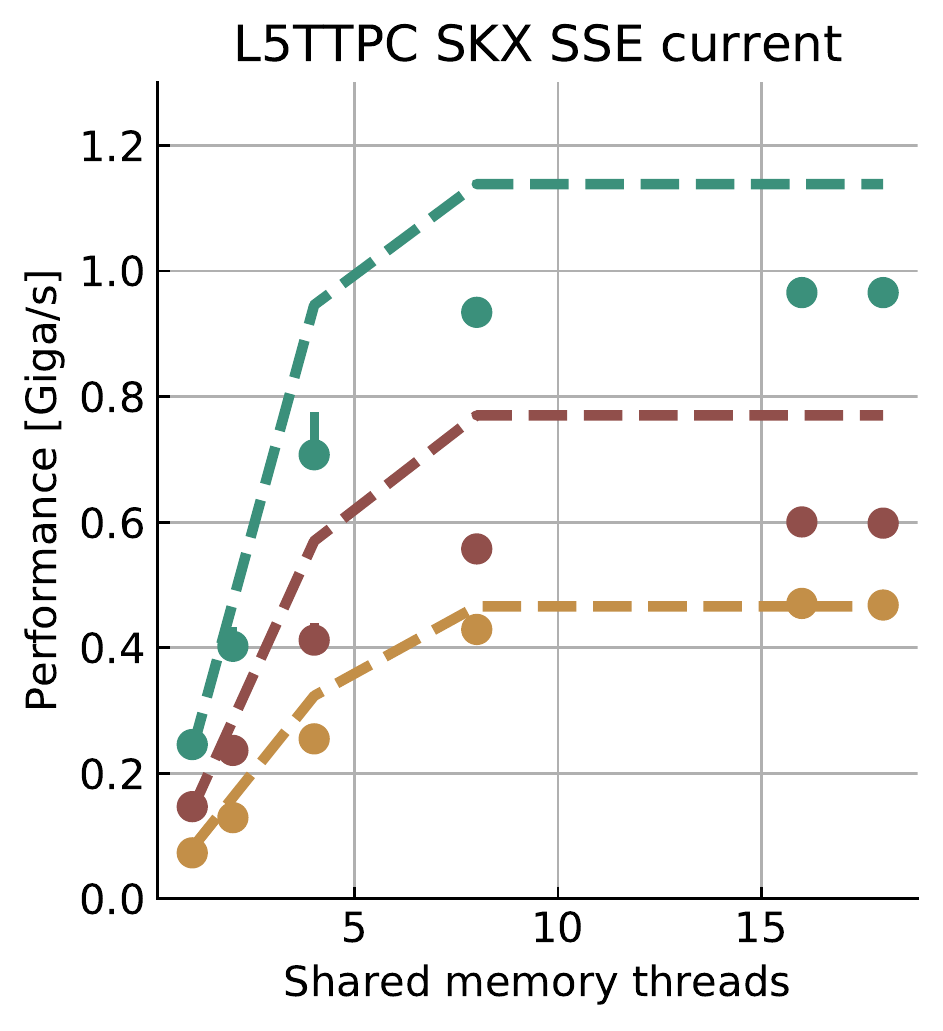}~
    \includegraphics[width=0.195\textwidth]{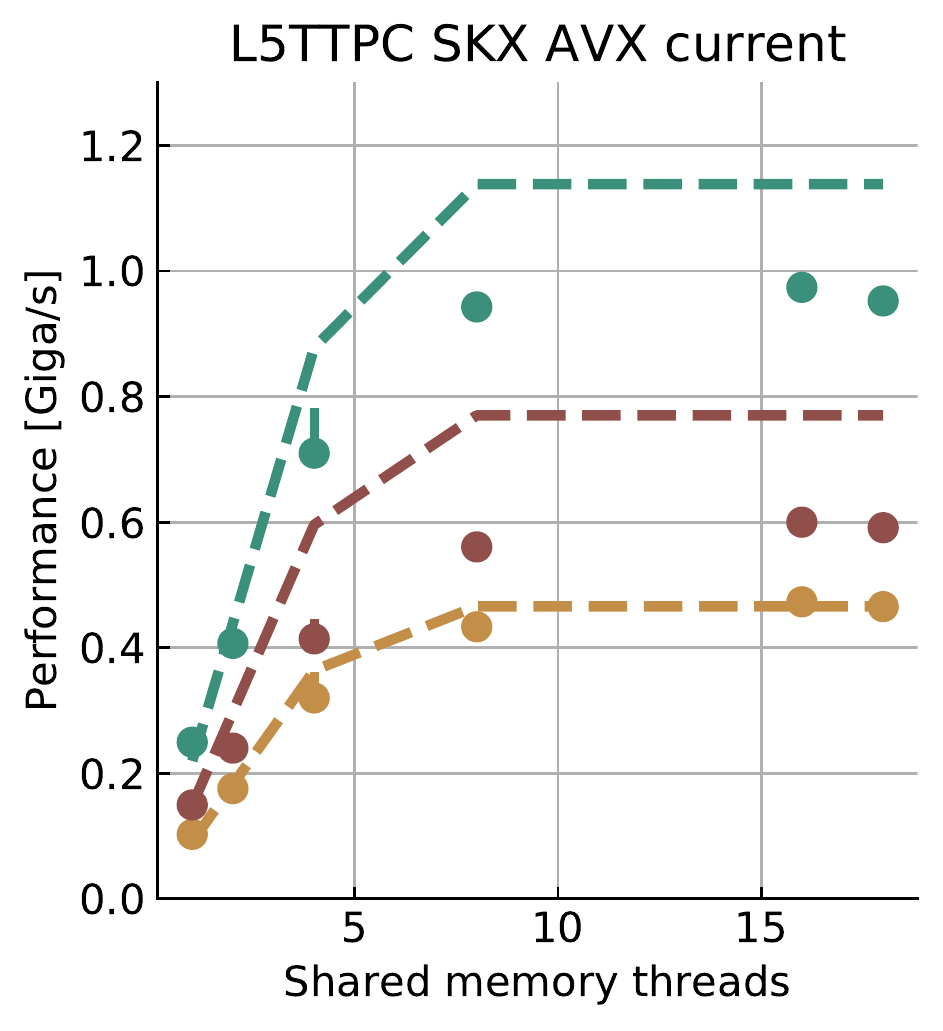}~
    \includegraphics[width=0.195\textwidth]{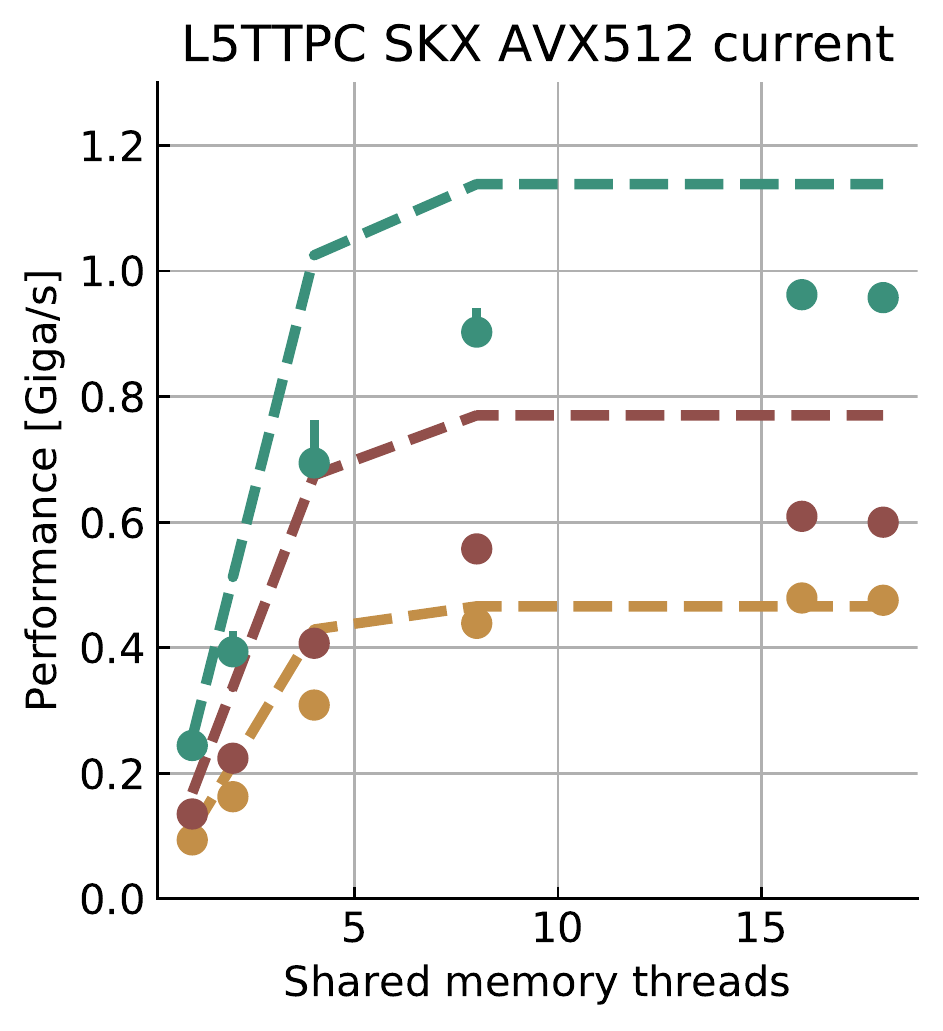}\\
    \includegraphics[width=0.195\textwidth]{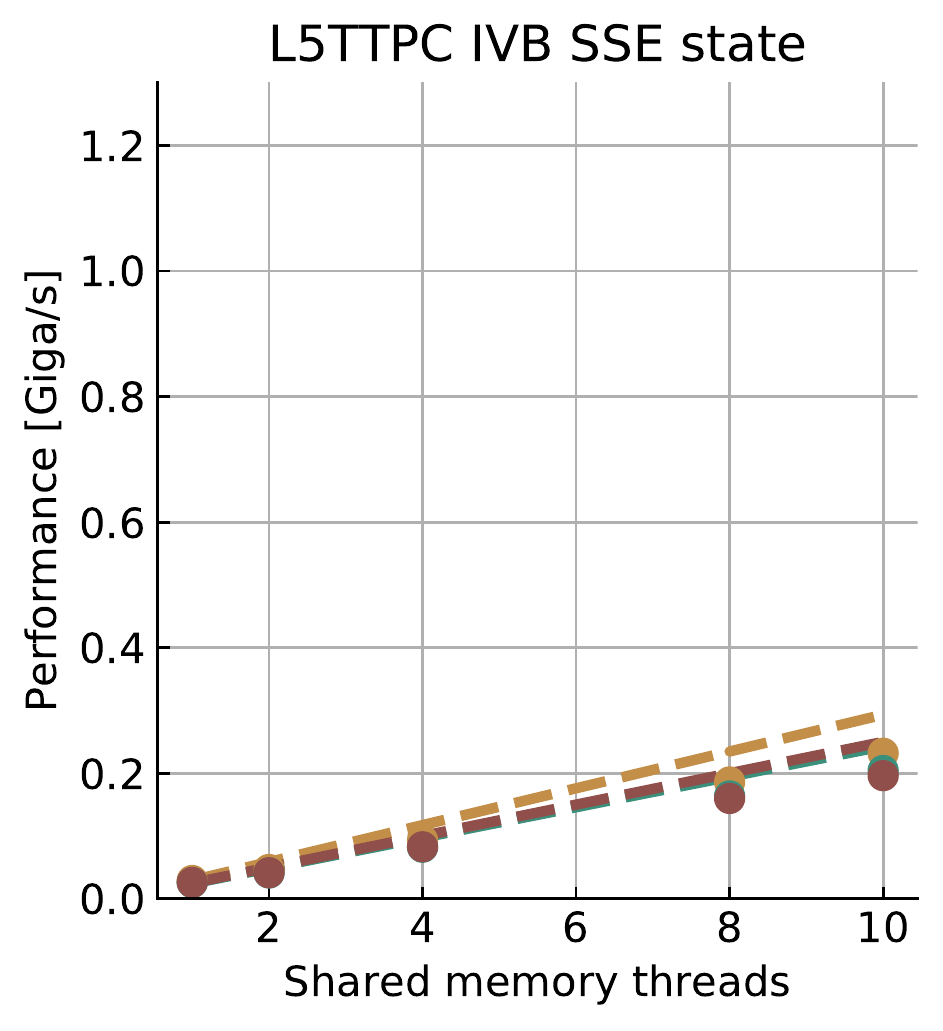}~
    \includegraphics[width=0.195\textwidth]{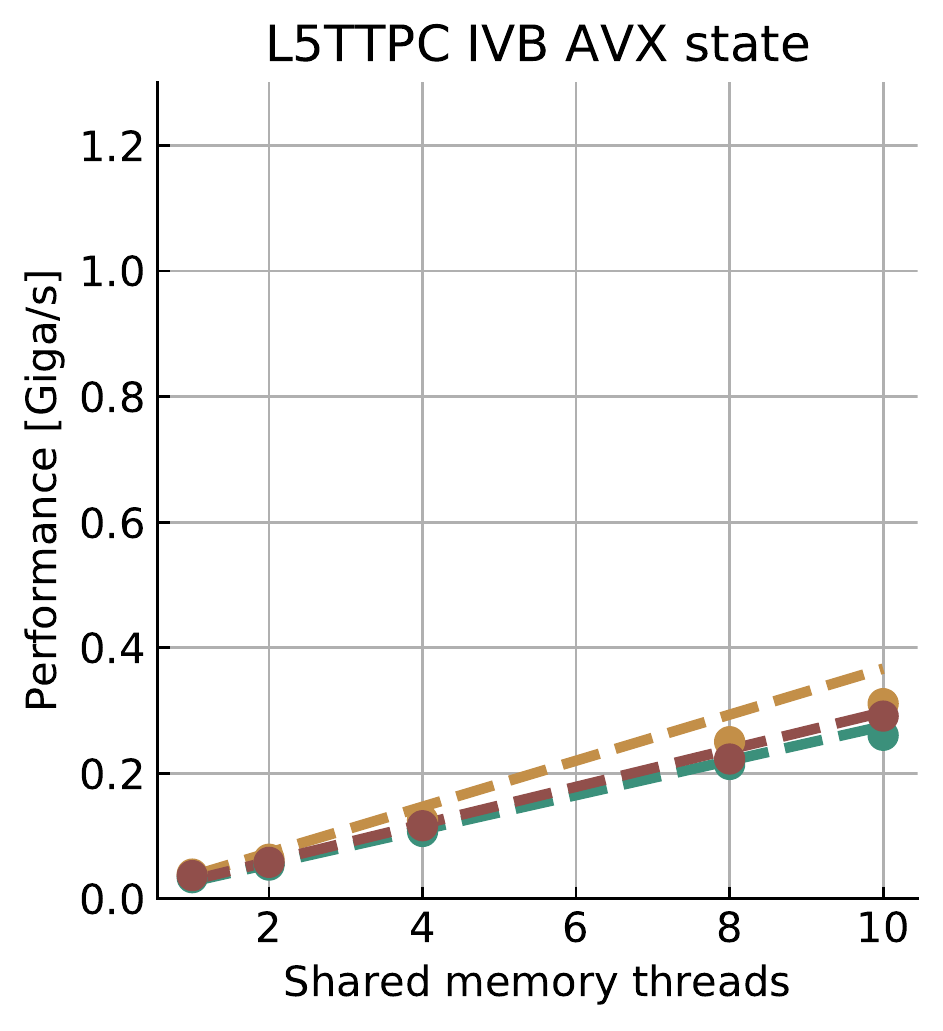}~
    \includegraphics[width=0.195\textwidth]{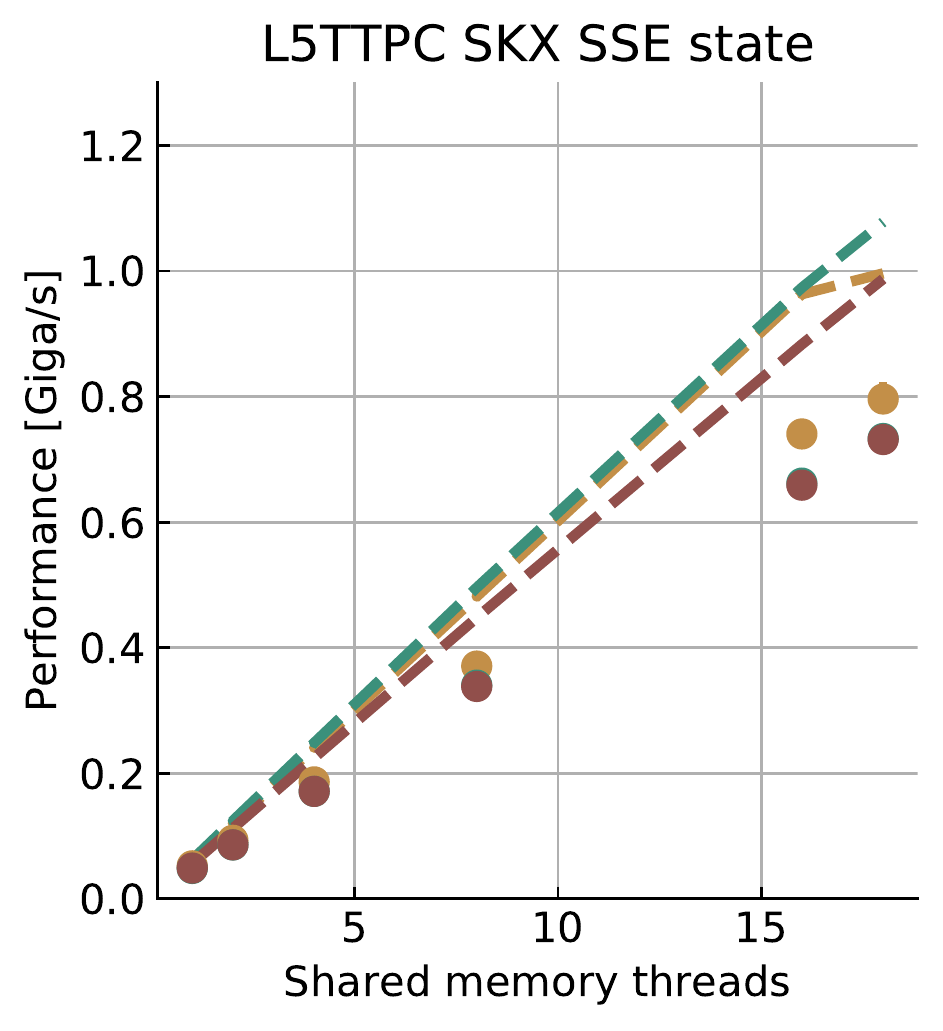}~
    \includegraphics[width=0.195\textwidth]{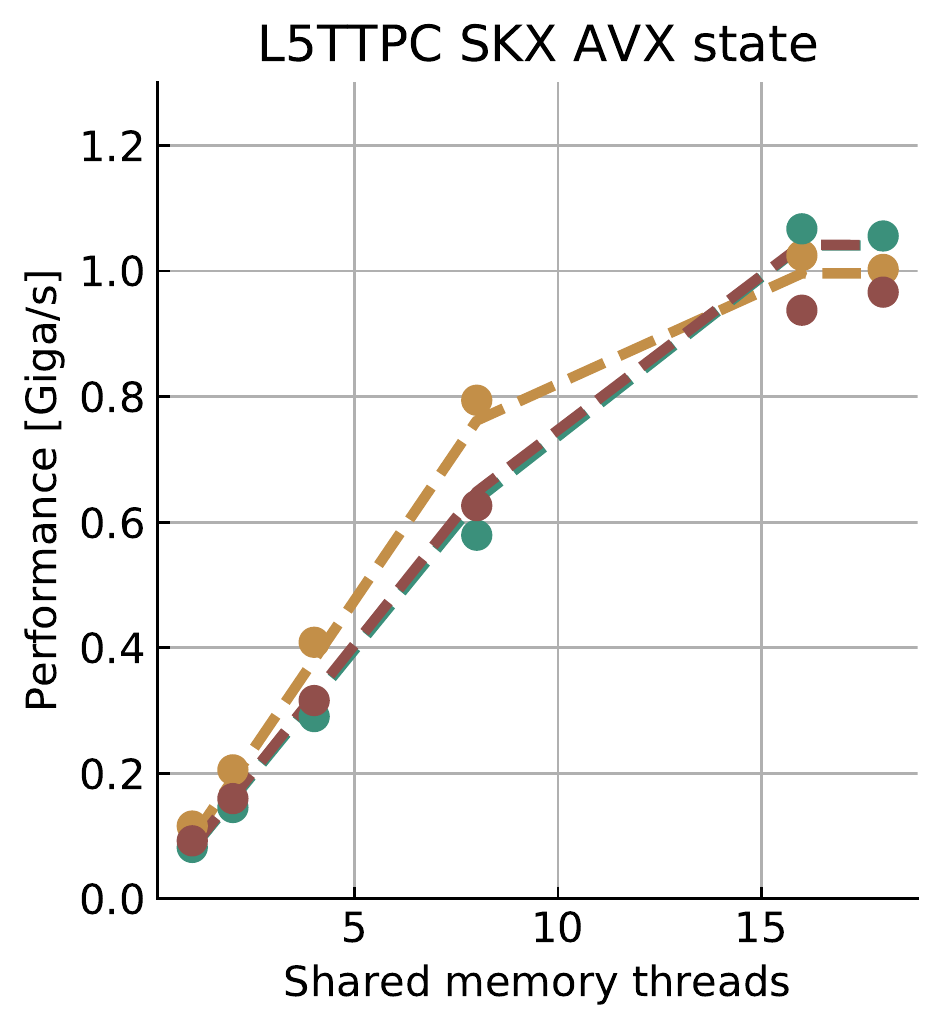}~
    \includegraphics[width=0.195\textwidth]{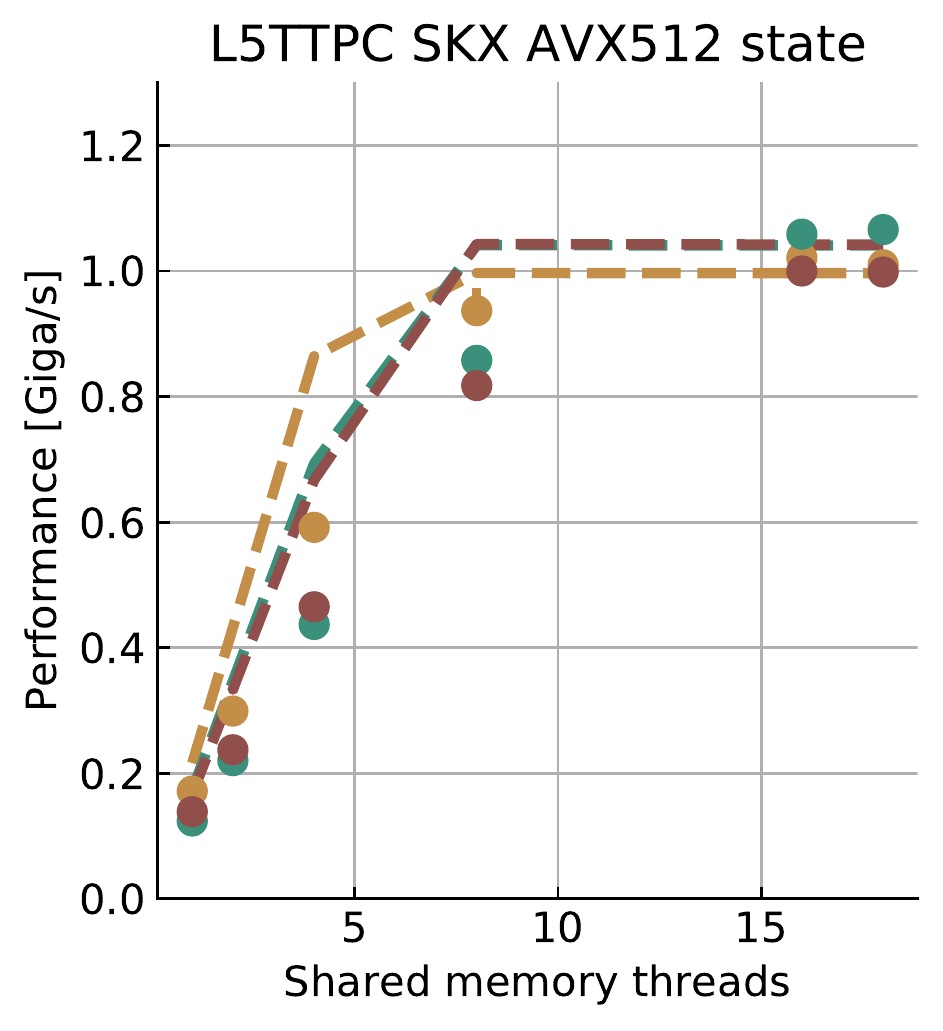}\\
    \includegraphics[width=0.2\textwidth]{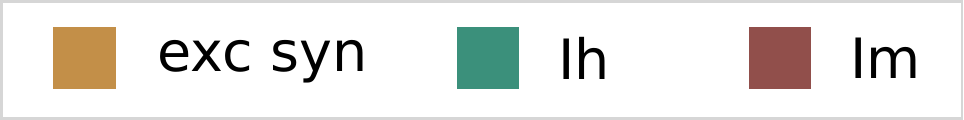} \\
    \caption{Performance predictions (dashed lines) and measurements (solid markers) for selected ion channel and state kernels, on all architectures and vectorization levels.
    Measurements points are computed as the median and errobars represent the 25- and 75-percentile out of 10 runs.
    Due to automatic clock frequency scaling, the performance predictions of each kernel were scaled by the kernel's average clock frequency to preserve consistency with the measurements.
    \emph{Current} kernels show a typical saturation behaviour at low thread counts while \emph{state} kernels either do not saturate at all (IVB), saturate at large thread counts (SKX-SSE, SKX-AVX) or saturate at moderate to low thread counts (SKX-AVX512).
    \label{fig_perf_scaling}}
\end{figure*}

Tables~\ref{tab_ivb} and~\ref{tab_skx} show the predicted and measured runtimes of current and state kernels for the two architectures, all vectorization levels and serial vs.\ full-socket parallel execution, while Figure~\ref{fig_perf_scaling} presents the performance scaling of these kernels across the cores of a chip. 
Overall we observe a good match between the predicted and observed runtimes: excluding a few exceptions our predictions always fall within 15\% of the observations, and we are able to correctly capture the previously observed phenomenon that \emph{current} kernels have a strongly saturating behavior, while \emph{state} kernels need more cores to saturate or do not saturate at all (such as on IVB, and on SKX with SSE code).
This corroborates our statements about optimization and co-design strategies: Boost in-core performance via reducing expensive operations (divides and exponentials), using wide SIMD cores and high clock speed for \emph{state}, and look for a fast memory hierarchy to reduce the data delay of \emph{current} kernels.
As the runtime of state and current kernels decreases, we expect the relative importance of spike delivery and linear algebra to increase. We will cover these two kernels in Sections~\ref{sec_linalg} and~\ref{sec_delivery}.

In the rest of this section, we address some of the largest deviations between measurements and predictions by providing a tentative explanation for the failure of our performance model.
As stated in the state kernel Sections~\ref{sec_ion_state} and~\ref{sec_syn_state}, a long critical path in the loop kernel code could be weakening the accuracy of our predictions due to a failure of the full throughput assumption.
We believe that, in order to improve our predictions, a cycle-accurate simulation of the execution and in particular of the OoO engine would be needed, thus invalidating our requirement for a simple analytical model.
At large thread counts the predictions for current kernels are always within a reasonable error bound, while those for state kernels can be off by as much as 30\%.
The state kernels' performance is often in a transitional phase between saturation and core-boundedness even at large thread counts, where the ECM model in the form we use it here is known to perform poorly as shown in~\cite{sthw15}. We do not plan to employ the adaptive latency penalty method as described in~\cite{hofmann2018accuracy} to correct for this discrepancy, because it is not only a
modification of the machine model but also requires a parameter fit for every individual
loop kernel. We believe that this is an undesirable trait in an analytic model. 
\begin{table*}
  \caption{Runtime for all current and state kernels on the IVB architecture
    (in-memory working set).
    Benchmark data is written as median $\pm$ interquantile range over ten runs.
    Both predicted and benchmark data is given in cycles per iteration.\label{tab_ivb}
    }
    \small\sf\centering
    \begin{tabular}{lcccccccc}
\toprule
 & \multicolumn{4}{c}{SSE} & \multicolumn{4}{c}{AVX} \\
    \cmidrule(lr){2-5} \cmidrule(lr){6-9}
& \multicolumn{2}{c}{serial} & \multicolumn{2}{c}{full socket} & \multicolumn{2}{c}{serial} & \multicolumn{2}{c}{full socket} \\
    \cmidrule(lr){2-3} \cmidrule(lr){4-5} \cmidrule(lr){6-7} \cmidrule(lr){8-9}
    kernel & pred & bench & pred & bench & pred & bench & pred & bench \\
\midrule
 exc syn current & 33.9 & 35.2$\pm$0.2 & 11.3 & 11.4$\pm$0.0 	& 31.9 & 28.9$\pm$0.9 & 11.3 & 11.4$\pm$0.1 \\
 inh syn current & 28.3 & 26.5$\pm$0.2 & 10.0 & 10.1$\pm$0.1 	& 27.3 & 26.4$\pm$0.2 & 10.0 & 10.1$\pm$0.1 \\
 NaTs2\_t current & 23.4 & 21.3$\pm$0.2 & 8.1 & 8.4$\pm$0.2 	& 28.0 & 21.0$\pm$0.2 & 8.1 & 8.2$\pm$0.2 \\
 Ih current & 13.3 & 12.0$\pm$0.0 & 5.1 & 5.0$\pm$0.1 	& 13.8 & 11.9$\pm$0.0 & 5.1 & 4.9$\pm$0.1 \\
 Im current & 21.5 & 19.0$\pm$0.2 & 7.5 & 7.9$\pm$0.1 	& 21.6 & 19.0$\pm$0.1 & 7.5 & 7.7$\pm$0.1 \\
 SKv3\_1 current & 22.0 & 19.9$\pm$0.1 & 7.7 & 7.9$\pm$0.2 	& 22.1 & 19.7$\pm$0.1 & 7.7 & 7.7$\pm$0.0 \\
 exc syn state & 75.0 & 75.4$\pm$0.0 & 7.5 & 9.5$\pm$0.0 	& 60.0 & 55.9$\pm$0.0 & 6.0 & 7.1$\pm$0.0 \\
 inh syn state & 75.0 & 73.5$\pm$0.1 & 7.5 & 9.3$\pm$0.0 	& 60.0 & 51.7$\pm$0.0 & 6.0 & 6.5$\pm$0.0 \\
 NaTs2\_t state & 220.5 & 162.7$\pm$2.1 & 22.0 & 20.4$\pm$0.0 	& 196.0 & 142.5$\pm$0.3 & 19.6 & 17.9$\pm$0.0 \\
 Ih state & 90.5 & 85.6$\pm$0.0 & 9.1 & 10.8$\pm$0.0 	& 80.0 & 65.5$\pm$0.0 & 8.0 & 8.4$\pm$0.0 \\
 Im state & 88.0 & 84.1$\pm$0.2 & 8.8 & 11.2$\pm$0.0 	& 74.0 & 59.6$\pm$0.6 & 7.4 & 7.6$\pm$0.1 \\
 SKv3\_1 state & 83.5 & 79.8$\pm$0.0 & 8.3 & 9.9$\pm$0.0 	& 73.0 & 60.7$\pm$0.1 & 7.3 & 7.5$\pm$0.0 \\
\bottomrule
\end{tabular}

\end{table*}
\begin{table*}
  \caption{Runtime for all current and state kernels on the SKX architecture
    (in-memory working set).
    Benchmark data is written as median $\pm$ interquantile range  over ten runs.
    Both predicted and benchmark data is given in cycles per iteration.\label{tab_skx}
    }
    \small\sf\centering
    \begin{tabular}{lcccccccccccc}
\toprule
 & \multicolumn{4}{c}{SSE} & \multicolumn{4}{c}{AVX}  & \multicolumn{4}{c}{AVX512} \\
    \cmidrule(lr){2-5} \cmidrule(lr){6-9} \cmidrule(lr){10-13}
& \multicolumn{2}{c}{serial} & \multicolumn{2}{c}{full socket} & \multicolumn{2}{c}{serial} & \multicolumn{2}{c}{full socket} & \multicolumn{2}{c}{serial} & \multicolumn{2}{c}{full socket} \\
    \cmidrule(lr){2-3} \cmidrule(lr){4-5} \cmidrule(lr){6-7} \cmidrule(lr){8-9} \cmidrule(lr){10-11} \cmidrule(lr){12-13}
    kernel & pred & bench & pred & bench & pred & bench & pred & bench & pred & bench & pred & bench  \\
\hline
 exc syn current & 25.9 & 28.6$\pm$0.0 & 4.5 & 4.5$\pm$0.1 	& 23.0 & 20.4$\pm$2.3 & 4.5 & 4.5$\pm$0.1 	& 19.5 & 22.3$\pm$1.7 & 4.5 & 4.4$\pm$0.1 \\
 inh syn current & 21.6 & 22.5$\pm$3.0 & 4.0 & 4.8$\pm$0.0 	& 19.8 & 22.5$\pm$2.0 & 4.0 & 4.8$\pm$0.1 	& 16.6 & 23.4$\pm$0.6 & 4.0 & 4.7$\pm$0.1 \\
 NaTs2\_t current & 17.8 & 16.5$\pm$1.1 & 3.2 & 4.1$\pm$0.1 	& 17.2 & 16.2$\pm$1.1 & 3.2 & 4.0$\pm$0.1 	& 14.9 & 16.8$\pm$0.7 & 3.2 & 4.0$\pm$0.1 \\
 Ih current & 9.7 & 9.3$\pm$0.4 & 2.0 & 2.4$\pm$0.1 	& 10.5 & 9.2$\pm$0.4 & 2.0 & 2.4$\pm$0.1 	& 9.0 & 9.4$\pm$0.4 & 2.0 & 2.4$\pm$0.0 \\
 Im current & 16.1 & 15.6$\pm$0.8 & 3.0 & 3.8$\pm$0.1 	& 15.4 & 15.3$\pm$0.8 & 3.0 & 3.9$\pm$0.1 	& 13.6 & 17.0$\pm$0.6 & 3.0 & 3.8$\pm$0.1 \\
 SKv3\_1 current & 16.5 & 14.9$\pm$0.7 & 3.1 & 3.8$\pm$0.1 	& 16.8 & 14.7$\pm$0.8 & 3.1 & 3.9$\pm$0.1 	& 14.0 & 15.4$\pm$0.4 & 3.1 & 3.8$\pm$0.1 \\
 exc syn state & 34.8 & 39.9$\pm$0.0 & 2.1 & 2.6$\pm$0.1 	& 22.0 & 18.1$\pm$0.1 & 2.1 & 2.1$\pm$0.1 	& 9.7 & 12.2$\pm$0.2 & 2.1 & 2.1$\pm$0.0 \\
 inh syn state & 34.8 & 40.2$\pm$0.0 & 2.1 & 2.6$\pm$0.0 	& 22.0 & 18.0$\pm$0.0 & 2.1 & 2.1$\pm$0.1 	& 9.7 & 12.2$\pm$0.3 & 2.1 & 2.1$\pm$0.1 \\
 NaTs2\_t state & 86.7 & 94.5$\pm$0.0 & 4.8 & 6.0$\pm$0.0 	& 64.5 & 51.1$\pm$0.0 & 3.8 & 4.0$\pm$0.1 	& 25.3 & 29.0$\pm$0.1 & 3.8 & 3.8$\pm$0.1 \\
 Ih state & 36.1 & 46.5$\pm$0.0 & 2.0 & 3.0$\pm$0.0 	& 26.4 & 25.6$\pm$0.0 & 2.0 & 2.0$\pm$0.0 	& 12.1 & 16.9$\pm$0.1 & 2.0 & 2.0$\pm$0.0 \\
 Im state & 38.6 & 44.3$\pm$0.1 & 2.1 & 2.9$\pm$0.0 	& 25.9 & 22.7$\pm$0.1 & 2.0 & 2.2$\pm$0.0 	& 12.6 & 15.1$\pm$0.3 & 2.0 & 2.1$\pm$0.0 \\
 SKv3\_1 state & 34.0 & 40.8$\pm$0.0 & 1.9 & 2.7$\pm$0.0 	& 24.5 & 21.7$\pm$0.0 & 1.4 & 1.6$\pm$0.0 	& 16.1 & 13.3$\pm$0.1 & 1.3 & 1.5$\pm$0.0 \\
\bottomrule
\end{tabular}

\end{table*}
\begin{table}
    \caption{Predicted and measured data volume per iteration from main memory.
    Benchmark data is written as median $\pm$ interquantile range over five runs, all vectorization levels and all thread counts.\label{tab_mem} }
    \small\sf\centering
    \begin{tabular}{lccc}
\toprule
 kernel          &   pred [B] & IVB meas [B]   & SKX meas [B]   \\
\midrule
 exc syn current  &      205.3 & 205.2$\pm$2.8 &  207.1$\pm$2.1\\
 inh syn current  &      181.3 & 183.3$\pm$5.2 &  204.0$\pm$8.4\\
 NaTs2\_t current &      148.0 & 144.3$\pm$8.2 &  139.4$\pm$11.0\\
 Ih current       &       92.0 & 79.2$\pm$4.3  &  80.2$\pm$9.2 \\
 Im current       &      136.0 & 128.9$\pm$5.8 &  133.4$\pm$10.8\\
 SKv3\_1 current  &      140.0 & 128.8$\pm$8.0 &  128.1$\pm$13.3\\
 exc syn state    &       96.0 & 95.6$\pm$1.9  &  94.3$\pm$1.3 \\
 inh syn state    &       96.0 & 91.3$\pm$5.3  &  88.6$\pm$4.6 \\
 NaTs2\_t state   &      172.0 & 197.4$\pm$1.9 &  166.2$\pm$2.2\\
 Ih state         &       92.0 & 88.0$\pm$0.3  &  87.7$\pm$1.2 \\
 Im state         &       92.0 & 118.0$\pm$5.8 &  89.1$\pm$2.0 \\
 SKv3\_1 state    &       60.0 & 92.5$\pm$8.3  &  56.6$\pm$2.1 \\
 linear algebra   &       88.0 & 90.6$\pm$7.6  &  90.7$\pm$4.2 \\
\bottomrule
\end{tabular}

\end{table}
\subsection{Special kernels: linear algebra}\label{sec_linalg}
The most common approach for time integration of morphologically detailed neurons is to use an implicit method (typically backward-Euler or Crank-Nicolson) in order to take advantage of its stability properties for stiff problems.
In~\cite{hines1984efficient} a linear-complexity algorithm based on~\cite{thomas1949elliptic} was introduced to solve the quasi-tridiagonal system arising from the branched morphologies of neurons.
This algorithm is based on a sparse representation of the matrix using three arrays of values (\lstinline!vec_a,vec_b,vec_d! representing the upper, lower and diagonal of the matrix, respectively) and one array of indices (\lstinline!parent_index!). It is structured in two main phases: triangularization and a backward substitution. The code is shown in Listing~\ref{lst_linalg}. The boundary loop in the middle is executed but its trip count is so short in practice that we can ignore it in the analysis.
\begin{lstlisting}[language=C, tabsize=2, basicstyle=\footnotesize\ttfamily,float,
caption={Linear algebra kernel},label={lst_linalg}]
//triangularization
for (i = ncompartments - 1; i >= ncells; --i) {
  p = vec_a[i]/vec_d[i];
  vec_d[parent_index[i]] -= p*vec_b[i];
  vec_rhs[parent_index[i]] -= p*vec_rhs[i];
}
//solve boundaries (ignored)
for (i = 0; i < ncells; ++i) {
  vec_rhs[i] /= vec_d[i];
}
//backward substitution
for (i = ncells; i < ncompartments; ++i) {
  vec_rhs[i] -= vec_b[i]*vec_rhs[parent_index[i]];
  vec_rhs[i] /= vec_d[i];
}
\end{lstlisting}

To construct a performance model for this kernel we must tackle a few challenges: Indirect accesses make it difficult to estimate the data traffic, and dependencies between loop iterations could break the full-throughput hypothesis.
Moreover, a yet-unpublished optimized variant of the algorithm proposed in~\cite{hines1984efficient} that exploits a permutation of node indices to maximize data locality is executed by default by the simulation engine\endnote{See the open-source code at \url{https://github.com/BlueBrain/CoreNeuron}. This permutation of node indices can be disabled with the command line argument \texttt{--cell-permute 0}.}.
For reasons of brevity of exposition we restrict our analysis to this optimized variant of the solver.
Additionally we will ignore the \lstinline!solve boundaries! loop in our analysis because its impact on the overall performance is always neglectable, for two reasons: the number of cells is always much smaller than the number of compartments so this loop makes very few iterations compared to the others, and there are no data dependencies so this loop can be trivially vectorized.

In order to give a runtime estimate we examine two corner-case scenarios.
The first, optimistic scenario assumes that indirect accesses can exploit spatial data locality in caches and thus do not generate any additional memory traffic.
The combined data traffic requirements of triangularization and back-substitution then amount to reading one element each from four double arrays and two integer arrays, and writing one element each to three double arrays, i.e., 88\,\byte\ per iteration.
Considering the opposite extreme, it might happen that at every branching point the value of \lstinline!parent_index[i]! is so much smaller than \lstinline!i! that this generates an additional cache line of data traffic through the full memory hierarchy.
We call this the worst-case branching hypothesis, in which we adjust the memory traffic predictions by assuming that every section boundary, i.e., the location of a potential discontinuity in the \lstinline!parent_index! array, requires a full cache line transfer of which only one variable will constitute useful data.

Even though the dependencies between loop iterations could potentially break the full-throughput hypothesis, considering that compartment indices are by default internally rearranged to optimize data locality we still use the full throughput as a basis for our predictions.
It should be noted that indirect addressing and potential loop dependencies hinder vectorization.
IACA reports that the combined inverse throughput of triangularization and back substitution amounts to 28\,\cycles/it for IVB and 8.12\,\cycles/it for SKX, while $T_{nOL} = 6$\,\cycles/it for both architectures.
This leads to the runtime predictions in Table~\ref{tab_linalg}.
\begin{table}[!h]
    \centering
    \caption{ECM model and serial measurements per scalar iteration [cy] for the linear algebra kernel. Vectorization levels are not considered because indirect write accesses prevent vectorization.\label{tab_linalg}}
    \begin{tabular}{lccc}
        \toprule
        & contributions & $T_{ECM}^{Mem}$ & measured\\[0.5mm]
        \midrule
        IVB & \ecm{28.0}{6.0}{2.8}{2.8}{4.8}{} &  28.0 & $32.6 \pm 4.4$\\[0.5mm]
        SKX & \ecm{8.1}{6.0}{1.4}{4.0}{1.9}{} &  13.3 & $18.8 \pm 5.3$\\[0.5mm]
        \bottomrule
    \end{tabular}
\end{table}

\begin{figure}
\centering
\includegraphics[width=0.95\columnwidth]{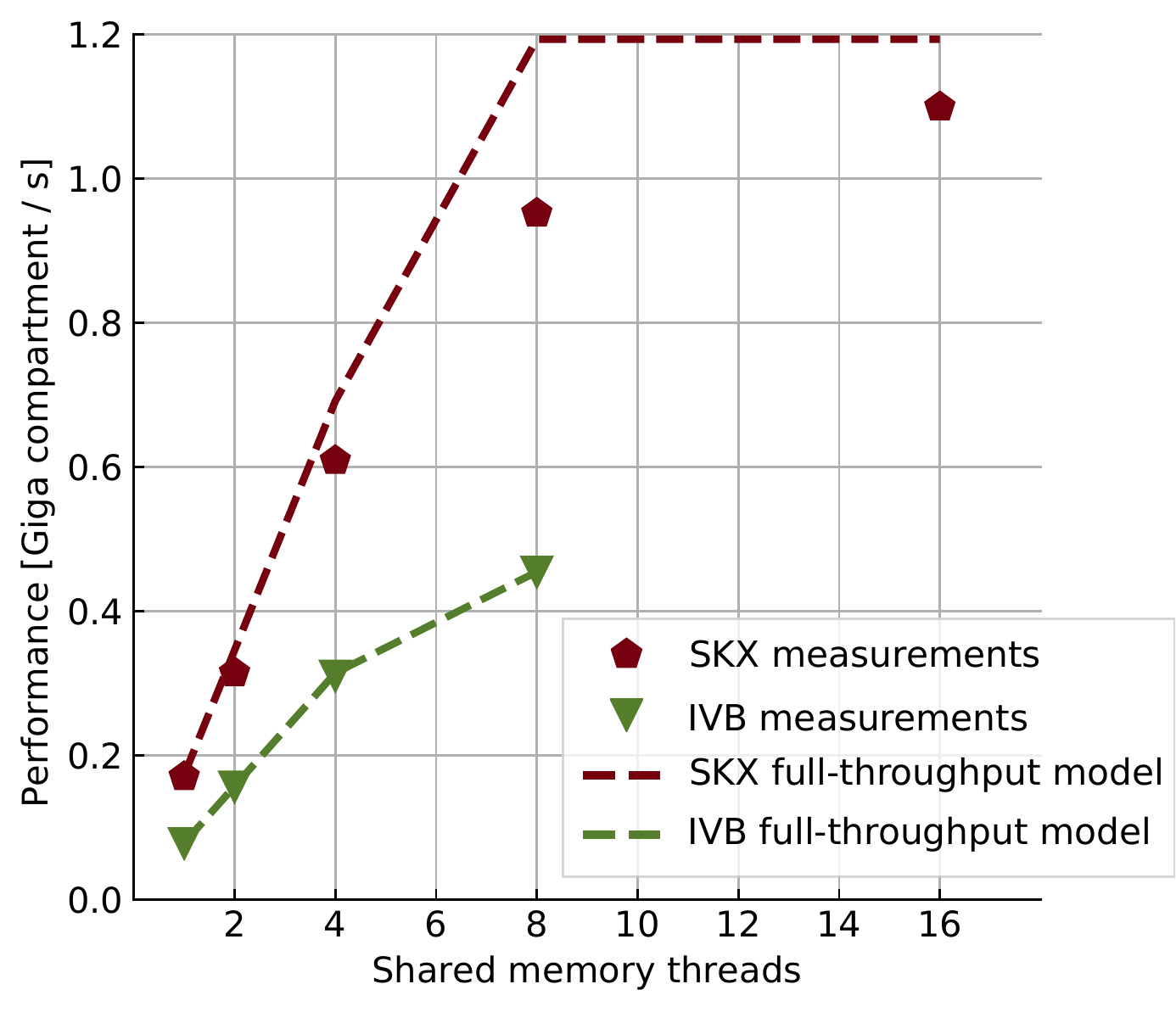}
\caption{Measured performance (markers) and predictions (lines) for the linear algebra kernel in Giga-compartments per second.
Dashed lines represent the model predictions in the optimistic full-throughput scenario. 
\label{fig_linalg}}
\end{figure}

We measured the performance of the linear algebra kernel on a specially designed dataset with a very large number of cells and neither ion channels nor synapses, thus ensuring that the only data locality effects are intrinsic to the algorithm and not a consequence of a small dataset.
Our predictions based on the full-throughput hypothesis are validated by measurements of both the performance (see Figure~\ref{fig_linalg}) and the memory traffic (last row in Table~\ref{tab_mem}).
This kernel highlights very strongly an important difference between the two architectures:
SKX has a much better divide unit, which is able to deliver one result every four cycles, whereas IVB's divider needs 14. This ratio is almost exactly reflected in the $T_{OL}$ prediction, although the triangularization kernel on SKX is actually load bound by a small margin.
This large difference in $T_{OL}$ causes different single-core bottlenecks: While the execution on IVB clearly core bound, it is strongly data bound on SKX\@.  The single-core medians are a little higher than predicted but also prone to some statistical variation; the best measured value is very close to the model. 
Saturation is predicted at six cores on IVB and seven cores on SKX. Starting from the newer architecture, the only way to boost performance would be to enhance the performance of the memory hierarchy (in serial mode) or the memory bandwidth (in parallel). Having more than ten cores per chip would be a waste of transistors.

We remark that it remains unclear whether the node permutation optimization is applicable in all cases or suffers from some constraints, and that our full-throughput predictions heavily rely on it.
Therefore it may happen that, in some cases where it is impossible to reorder the nodes effectively, our predictions would only provide an optimistic upper bound on performance.
\subsection{Special kernels: spike delivery}\label{sec_delivery}
Accounting for network connectivity and event-driven spike exchange between neurons is, in terms of algorithm design, the most distinguishing feature of neural tissue simulations.
In terms of performance, however, spike delivery plays a marginal role in the simulation of morphologically detailed neurons, rarely exceeding 10\% of the total runtime (see Figure~\ref{fig_breakdown}a).

The source code for the spike delivery kernel of AMPA/NMDA excitatory synapses is shown in Listing~\ref{lst_spkdel}\@.
For benchmarking purposes we executed this kernel as the body of a loop iterating over a vector of spike events, which was previously populated by popping a priority queue\endnote{See branch \lstinline!perf_eng_binq_bench! of \url{https://github.com/sharkovsky/CoreNeuron.git}}.
This only represents a small deviation from the original implementation in CoreNEURON, where the kernel is directly called at every pop of the priority queue.
However, it was necessary to implement this in order to separate the performance of the kernels from the performance of the queue operations.

\begin{lstlisting}[language=C, tabsize=2, basicstyle=\footnotesize\ttfamily,float,
    caption={Event-driven spike delivery kernel}, label={lst_spkdel}]
Event events[];
// loop over n spike_events
for(int e=0; e<n; ++e)
{
  Event spike_event = events[e];
  Target * target = spike_event.target;
  int weight_index = spike_event.weight_index;
  int type = target.type;
  int i = target.index;
  double _lweight_AMPA = _weights[weight_index];
  double _lweight_NMDA = _lweight_AMPA;
  _lweight_NMDA *= NMDA_ratio[i];
  _tsav[i] = t;
  u[i] = u[i] * exp(-(t-tsyn[i])/Fac[i] );
  u[i] += Use[i]*(1.-u[i]);
  R[i] = 1.-(1.-R[i])*exp(-(t-tsyn[i])/Dep[i]);
  Pr[i] = u[i]*R[i];
  R[i] = R[i] - u[i]*R[i] ;
  tsyn[i] = t ;
  A_AMPA[i] += Pr[i]*_lweight_AMPA*factor_AMPA[i];
  B_AMPA[i] += Pr[i]*_lweight_AMPA*factor_AMPA[i];
  A_NMDA[i] += Pr[i]*_lweight_NMDA*factor_NMDA[i];
  B_NMDA[i] += Pr[i]*_lweight_NMDA*factor_NMDA[i];
}
\end{lstlisting}
This kernel is characterized by erratic memory accesses indexed by \lstinline!i!, as well as several compute-intensive operations such as divisions and exponentials, thus making it challenging to model.
In terms of memory traffic we consider two scenarios: a best-case one in which all synapses are activated in memory-contiguous order and a worst-case scenario in which synapses are activated in random order.
In the best-case scenario we assume the execution engine will be able to fully pipeline the execution and hide all latencies.
Thus we base our performance predictions on either the full-throughput hypothesis or a critical path.
Given that this kernel requires a read-only transfer on seven double arrays, three integer arrays and one pointer array, and an update or write/write-allocate transfer on nine double arrays, we estimate a (best-case) memory traffic of 220\,\byte\ per iteration.
From IACA we learn that the inverse throughput of this kernel is 29.5\,\cycles/it on IVB and 27.6\,\cycles/it on SKX, while $T_{nOL}$ is 19.5\,\cycles/it on both architectures, and as with the linear algebra kernel the indirect accesses prevent vectorization.
Under the full-throughput assumption, this leads to the single-thread predictions per iteration shown in Table~\ref{tab_delivery}.

Given the complex chain of interdependencies in the kernel, we suspect that a CP effect could also be present.
For the IVB architecture we can directly use IACA with the \lstinline!-analysis LATENCY! option, while for SKX we resorted to using the estimate for the Haswell architecture (HSW) from IACA v2.1, because latency analysis is no longer supported in IACA v3.0.
The CP values are also reported in Table~\ref{tab_delivery}.
\begin{table}[!h]
    \centering
    \caption{ECM model per scalar iteration [cy] for the spike delivery kernel. Vectorization levels are not considered because indirect accesses prevent vectorization. On SKX the CP prediction is actually for the Haswell architecture (see text for details). \label{ta_delivery}}
    \begin{tabular}{lccc}
        \toprule
        & contributions & $T_{ECM}^{Mem}$ & CP \\[0.5mm]
        \midrule
        IVB & \ecm{85.1}{19.5}{6.9}{6.9}{12.1}{} &  85.1 & 207.0\\[0.5mm]
        SKX & \ecm{57.8}{19.5}{3.4}{9.2}{4.8}{} &  57.8 & 123.4\\[0.5mm]
        \bottomrule
    \end{tabular}
\end{table}

In the worst-case scenario we assume that a full cache line of data needs to be brought in from memory \emph{for every data access}.
Assuming that the variables \lstinline!spike_event.target! and \lstinline!spike_event.weight_index! can still be read contiguously, the kernel requires 27 noncontiguous data accesses plus reading from one pointer and one integer array, which amounts to a predicted memory traffic of 1740\,\byte\ per iteration.
Estimating the runtime is more complex: On the one hand, it seems clear that the memory requests to arbitrary locations should have an effect on performance. On the other hand, this kernel does not have the typical latency-bound structure in which an iteration requires the full completion of the previous one before being executed.
Indeed, multiplying the number of memory accesses by the memory latency leads to a prediction that is more than ten times too pessimistic.
Instead, we created a synthetic stream-copy benchmark with a similar number of memory accesses and the same access pattern and determined the average latency per memory access to be around $20.1 \pm 1.3$ cycles for both architectures.
To obtain a runtime prediction for the serial execution we then multiply this average latency by the number of memory accesses, yielding a prediction of 540\,\cycles/it.
To extend this to the multi-threaded case, we assume that either the bandwidth is saturated (and thus our performance prediction corresponds to the \Rlm) or the performance scales linearly with the number of threads.

\begin{figure*}
\centering
\includegraphics[width=0.75\textwidth]{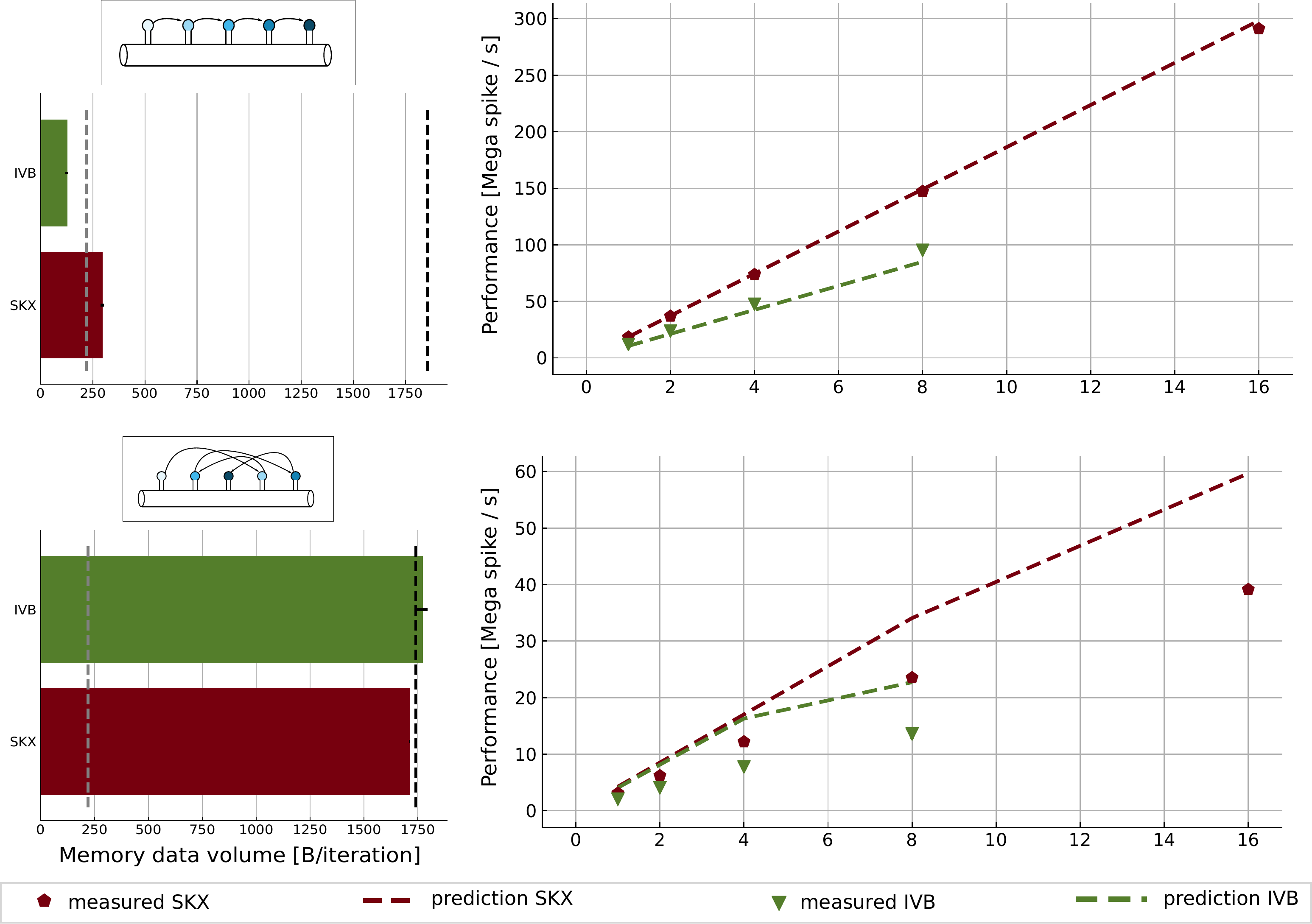}
\caption{ECM model and measurements for the spike delivery kernel.
    \emph{Top:} best-case scenario where synapses are activated in contiguous memory order.
    In this case there is no excess data traffic as shown by the bar plot on the left.
    The performance predictions on the right (dashed lines) are made by assuming that the kernel's runtime is equal to its critical path as predicted by IACA.
    Measurements (solid markers) substantiate this hypothesis.
    \emph{Bottom:} worst-case scenario where synapses are activated in random order.
    This scenario corresponds to the typical use-case.
    We assume that for every array access a full cache line of data traffic is generated, but only one element of the array is relevant.
    \label{fig_delivery}}
\end{figure*}

The validation of the model is shown in Figure~\ref{fig_delivery} and Table~\ref{tab_delivery}.
In the serial best-case scenario the measured runtime is so close to the CP-assumption that we can safely discard the full-throughput hypothesis and assume that under ideal memory access conditions this kernel is bounded by the dependencies within one loop iteration.
In the worst-case scenario, while the data volume predictions are quite correct, the runtime predictions are off by factors from 50\% up to 100\% (see also Table~\ref{tab_delivery}).
A reason for this could be that a CP estimate should be added to the memory access latency.
Unfortunately this does not give a sufficiently convincing improvement in the estimates: On IVB, IACA computes a CP of 79\,\cycles/it to which we should add twice the latency of a scalar exponential, benchmarked to be around 64\,\cycles.
This leads to an adjusted prediction of 747\,\cycles/it, which is still far from the measured 1087\,\cycles/it.
Considering that only a strikingly correct prediction would justify an adjustment to our model, we prefer to keep the old but simpler estimate. One should add that the worst-case scenario is beyond the applicability of the ECM model, so our analysis stretches the model very far. 
\begin{table}
    \centering
    \caption{Spike delivery runtime predictions and median measurements ($\pm$ interquantile range) under the best-case (BC) and worst-case (WC) scenarios, in serial (S) and parallel (P) execution.
    In the case of parallel execution we report the values for 8 threads on IVB and 16 threads on SKX.
    \label{tab_delivery}}
    \begin{tabular}{lcccc}
        \toprule
         & \multicolumn{2}{c}{Runtime IVB [cy]} & \multicolumn{2}{c}{Runtime SKX [cy]} \\
         \cmidrule(lr){2-3} \cmidrule(lr){4-5}
         & pred  & meas & pred  & meas \\
        \midrule
        BC-S  & 207.0 & $183.9  \pm 0.5$  & 123.4  & $122.1 \pm 0.5$ \\
        WC-S  & 540.0 & $1064.8 \pm 55.6$ & 540.0  & $740.0 \pm 2.1$ \\
        BC-P  & 25.9  & $23.1   \pm 0.0$  & 7.7    & $7.9   \pm 0.1$ \\
        WC-P  & 96.8  & $161.7  \pm 11.3$ & 45.0   & $58.8  \pm 0.1$ \\
        \bottomrule
    \end{tabular}
\end{table}
\section{Discussion}\label{sec:discussion}

Using the ECM performance model we have analyzed the performance profile of the simulation algorithm of morphologically detailed neurons as implemented in the CoreNEURON package.
Within its design space, the ECM model yielded accurate predictions for the runtime of 13 kernels on real-world datasets. It must be stressed that
some of these kernels are rather intricate, with hundreds of machine instructions and many parallel data streams. This confirms that analytic modeling is good for more than simple, educational benchmark cases. We have also, for the first time, set up the ECM model for the Intel Skylake-X architecture, whose cache
hierarchy differs considerably from earlier Intel server CPUs. Our analysis shows
that the non-overlapping assumption applies there as well, including all data
paths between main memory, the L2 cache and the victim L3. Note that a
reproducibility appendix is available at~\cite{BBP-ECM-RA}\@.

As expected, the modeling error was larger in situations where the bottleneck was neither streaming data access nor in-core instruction throughput. 
By making a few simplifying assumptions we were still able to predict with good accuracy the performance of a kernel with a complex memory access pattern and dependencies between loop iterations such as the tridiagonal Hines solver~\cite{hines1984efficient}.

On the other hand, if the bottleneck is the memory latency, which is the case with the  spike delivery kernel, the ECM model could only provide upper and lower bounds.
In this case where the deviation from the measurement was especially large, we could at least pinpoint possible causes for the failure. It is left to future work on the foundations of the ECM model to extend its validity in those settings.

In conclusion, the ECM model was always able to correctly identify the computational characteristics and thus the bottlenecks of the 14 kernels under investigation, thus providing valuable insight to the performance-aware developers and modelers.
In the following we use these crucial insights to give clear guidelines for both the optimization of simulation code and the co-design of an ``ideal'' processor architecture for neuron simulations. We mostly concentrate on the Skylake-X architecture since it is the most recent one, and only discuss results for Ivy Bridge where necessary.

\begin{table}
\caption{Possible causes for degradation of accuracy in ECM model.\label{tab_fails}}
\begin{tabular}{lcc}
\toprule
prediction is \dots & data-bound kernel & core-bound kernel \\
\midrule
optimistic  & memory latency & critical path \\
pessimistic & data locality  & OoO engine \\
\bottomrule
\end{tabular}
\end{table}

\subsection{Small networks (in cache)}

\paragraph{Serial performance properties of small networks}\mbox{}

One of the main insights offered by the ECM model is the possibility to identify
and quantify the performance bottleneck of each kernel.  In the simulation of
morphologically detailed neurons, we found that \emph{ion channel current}
kernels are data bound while \emph{all state} kernels are
core bound for all cache levels, all SIMD levels and both architectures considered.
The case of \emph{excitatory synapse current}
kernel was special in that on both SKX and IVB, the kernel was core bound as
long as the dataset fits in the caches, but switched to data-bound when the
data comes from memory.  This effect was most prominent on SKX-AVX512.  In the
extreme strong scaling scenario where data fits in the cache, this points to two optimizations: optimize expensive operations such as \lstinline!div! and \lstinline!exp!
for \emph{all state} kernels and the \emph{excitatory synapse current} kernel, and minimize data movement for the \emph{ion channel current} kernels.
In terms of co-design, high-frequency cores with high-throughput instructions are ideal for \emph{all state} kernels while fast data-paths within the cache hierarchy would optimize \emph{ion channel current} kernels.

\paragraph{SIMD parallelism and small networks}\mbox{}

The possibility to execute high-throughput SIMD vector instructions
can potentially provide great returns in terms of speedup at a low
hardware and programming cost.  In this analysis we observed that
wider SIMD units were indeed capable of providing benefits in terms of
reduced runtime, but we also failed to observe the ideal speedup
factor.  Moreover, Skylake-X showed diminishing returns as the SIMD
units grew wider.  Applying the ECM model to the scenario where data
comes from cache we discovered that \emph{all state} kernels show
significant speedups from vectorization, and would benefit even more
from even wider SIMD units.  The \emph{synapse current} kernels
benefit from SIMD instructions at least for data in the L1 or L2
cache.  \emph{Ion channel current} kernels show only small speedups
from vectorization because their performance is solely determined by
the speed of the data transfer, even when the working set fits into a
cache.

The importance of high-throughput \lstinline!exp! and
\lstinline!div! functions cannot be overrated, which is punctuated by the large performance
gain from Ivy Bridge to Skylake-X for kernels where these functions
contribute significantly to the runtime. We have 
observed that the compiler was sometimes not able to eliminate
expensive divide operations, although this was possible and allowed by
the optimization flags.

\subsection{Large networks (out of cache)}


\paragraph{Memory bandwidth saturation of large networks}\mbox{}

\begin{figure}
    \includegraphics[width=0.99\columnwidth]{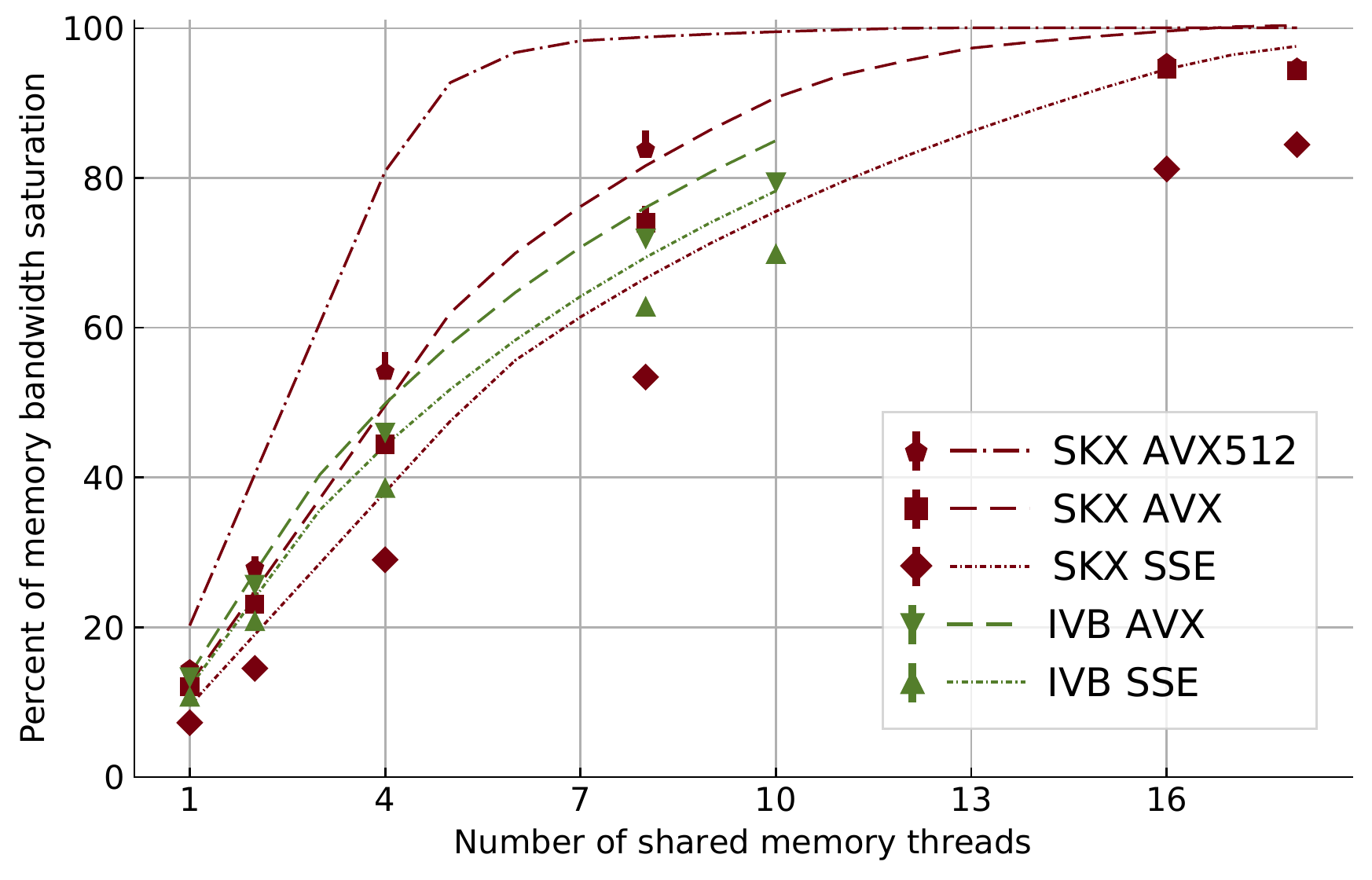}
    \caption{Predicted and measured memory bandwidth utilization, as a fraction of the maximum memory bandwidth.
      Dashed lines are obtained by predicting the average memory bandwidth
      of the full application while solid markers represent bandwidth
      measurements by likwid.
    Due to automatic clock frequency scaling, the maximum memory bandwidth was rescaled by the ratio of average clock frequency to nominal clock frequency.
      On SKX, AVX512 code can saturate the memory bandwidth of the socket at
      less than half the total number of cores even at the base clock
      frequency.
    \label{fig_satur}}
\end{figure}

At constant memory bandwidth, a sufficient number of cores and/or high enough
clock speed will render almost every code memory bound. One of the key insights
delivered by the ECM model is how many cores are required to achieve saturation
of the memory bandwidth during shared-memory execution, and what factors this
number depends on.  We applied saturation analysis to the full simulation loop
by predicting the memory bandwidth of each kernel for different numbers of
cores and compared it to the ratio of measured memory bandwidth to theoretical
maximum bandwidth, weighted by the runtime of each kernel.
Figure~\ref{fig_satur} shows the results, highlighting the remarkable power of
the AVX512 technology on SKX, which is able to almost fully saturate the memory
bandwidth using only seven cores.  Since in-core features come essentially for
free but more cores are more expensive, this means that in the max-filling
scenario where the number of neurons being simulated is large and the data fits
in main memory, the most cost-effective hardware platform for this code among
the architectures considered is one with AVX512 support, high clock speed and a
moderate core count.  To further quantify the tradeoff between clock speed and
saturation on SKX-AVX512 we computed the saturation point, which we define as
the number of threads required to utilize at least 90\% of the theoretical
memory bandwidth, at different clock frequencies for the SKX architecture
(assuming no clock frequency reduction).  The results in
Table~\ref{tab_satur_freq} highlight once again that, as long as the working
set is in main memory, vectorization
pushes the bottleneck towards the memory bandwidth in the shared memory
execution. We have to allow some room for error in the measurements of the
memory bandwidth and the over-optimistic ECM model near the saturation point as
shown in \cite{sthw15}, but the model indicates clearly that cores can be
traded for clock speed, which provides a convenient price-performance
optimization space. 
\begin{table}
    \caption{Saturation point as predicted by the ECM model as a function of clock frequency. The saturation point is here defined as the number of cores required to reach 90\% of the maximum memory bandwidth utilization. For modeling purposes we consider the ideal case where there is no clock frequency capping for large vector registers.\label{tab_satur_freq}}
    \begin{tabular}{lcc}
        \toprule
        & CPU @ 2.3 GHz & CPU @ 3.5 GHz \\
        \midrule
         SKX SSE  & 16 & 11 \\
         SKX AVX  & 12 & 8 \\
         SKX AVX512 & 6 & 4 \\
         \bottomrule
    \end{tabular}
\end{table}
\paragraph{Wide SIMD and large networks}\mbox{}

For in-memory data sets, wide SIMD execution helps to push the
saturation point to a smaller number of cores, as shown in
Table~\ref{tab_satur_freq} and Figure~\ref{fig_satur}, but
it will certainly not increase the saturated performance.
Hypothetical hardware with even wider SIMD units
would thus have to be supported by a larger memory
bandwidth to be fully effective. Moreover, as clearly
shown by the ECM model analysis, wider SIMD execution
would ultimately make even the \emph{state} kernels
data bound. In the mid-term future it would hence be
advisable to put more emphasis on fast clock speeds and
better memory bandwidth than on pushing towards wider
SIMD units, at least for the workloads discussed in this work. 

When choosing the most fitting cluster
architecture one is thus left with the decision between a larger
number of high-frequency chips with moderate memory bandwidth and a
smaller amount of lower-frequency chips with large memory bandwidth and
more cores. Roughly speaking, larger bandwidth is more expensive than
faster clock speed, but the decision has to be made according to
the market and pricing situation at hand, which  unfortunately
tends to be rather volatile.

\paragraph{Memory hierarchy for large networks}\mbox{}

There is practically no temporal locality in
the data access patterns of almost all kernels. This means that cache size
is insignificant for determining the performance of large networks of detailed
neuron simulations. Unfortunately, cache size is not a hardware parameter
that one is free to choose when procuring clusters of
off-the-shelf CPUs. 
Moreover, using the decomposition of the runtime by the
ECM model we observe that contributions from different levels of the memory
hierarchy are rather evenly distributed.  Hence, the
runtime of data-bound kernels could best be improved by reducing the data
volume.
A common programming technique to solve this problem is \emph{loop fusion}, by which two
or more back-to-back kernels that read or write some common data structures are
cast into a single loop in order to leverage temporal locality and thus increase
the arithmetic intensity. The structure of the NEURON code does not
easily allow this.


\section{Conclusions}

In this work we have demonstrated the applicability of the ECM analytic performance model to analyze and predict the bottlenecks and runtime of simulations of biological neural networks.
The need for such modeling is demonstrated by the ongoing development efforts to optimize simulation code for current state of the art HPC platforms, coupled with demands for simulators able to handle faster and larger datasets on present and future architectures.
Using the performance model we identified high-frequency cores capable of high-throughput \lstinline!div! and \lstinline!exp! operations and wide cache data paths as the most desirable features for real-time simulations of small neuron networks, while high memory bandwidth, few cores with moderate to high SIMD parallelism and a shallow memory hierarchy are the ideal hardware characteristics for simulations of large networks. No attempts have been made so far towards porting NEURON kernels to traditional vector processors (which have again become available recently), and porting to GPGPUs is still in an exploratory phase, but at least for large networks, where abundant parallelism is available, the characteristics we have identified let us expect speedups according to the memory bandwidth difference to standard multicore CPUs: a device with 1\,\TBS\ of memory bandwidth, such as the SX-Aurora ``Tsubasa'' by~\cite{tsubasa}, should outperform one Skylake-X socket by a factor of 9--10\@.

In the reconstruction and simulation of brain tissue, performance engineering and modeling is now a pressing issue limiting the scale and speed at which computational neuroscientists can run \emph{in silico} experiments.
We believe that our work represents an important contribution in understanding the fundamental performance properties of brain simulations and preparing the community for the next generation of hardware architectures.

\paragraph{Future work}\mbox{}

Two shortcomings hinder the comprehensive applicability of the ECM
model for all the kernels in CoreNEURON: the inability to correctly
describe latency-bound data accesses, and long critical paths in the
loop body. Both shortcomings may be addressed by refining the model,
i.e., endowing it with more information about the processor
architecture. This data, however, is not readily available (and it
might never be).  In case of critical path analysis, the Open Source
Architecture Code Analyzer (OSACA) by~\cite{Laukemann:2018} is planned
to become a versatile substitute for IACA, which does not provide
critical path prediction for modern Intel CPUs. Data latency
support would require a fundamental modification of the model, and
work is ongoing in this direction.

From the point of view of the simulation algorithm and implementation, given
the delayed nature of the dependencies between neuron connections, a potentially
very effective optimization could be made by looping through the time steps
within a minimum network delay for each neuron, nested within a loop over all
neurons, thus potentially allowing the algorithm to exploit temporal locality
of data.  This optimization is already implemented in CoreNEURON but requires
to generate many datasets comprising at most a few neurons instead of a single
dataset with many neurons, so it was not considered in this study.  The ECM model
provides a way to assess the tradeoffs of this approach but its validation is
still a work in progress.

\section*{Acknowledgments}

This work has been funded by the EPFL Blue Brain Project (funded by the Swiss ETH board).
The authors would like to thank Johannes Hofmann for fruitful discussions about low-level benchmarking and Thomas Gruber for his contributions to the LIKWID framework. We are also indebted to the Blue Brain HPC team for helpful support and discussion regarding CoreNEURON.

\theendnotes
\bibliographystyle{SageH}
\bibliography{paper}

\begin{thebibliography}{38}
\providecommand{\natexlab}[1]{#1}
\providecommand{\url}[1]{\texttt{#1}}
\providecommand{\urlprefix}{URL }
\expandafter\ifx\csname urlstyle\endcsname\relax
  \providecommand{\doi}[1]{DOI:\discretionary{}{}{}#1}\else
  \providecommand{\doi}{DOI:\discretionary{}{}{}\begingroup
  \urlstyle{rm}\Url}\fi

\bibitem[{Ananthanarayanan and Modha(2007)}]{ananthanarayanan2007anatomy}
Ananthanarayanan R and Modha DS (2007) Anatomy of a cortical simulator.
\newblock In: \emph{Proceedings of the 2007 ACM/IEEE conference on
  Supercomputing}. ACM, p.~3.

\bibitem[{Bhalla(2012)}]{bhalla2012multi}
Bhalla US (2012) Multi-compartmental models of neurons.
\newblock In: \emph{Computational Systems Neurobiology}. Springer, pp.
  193--225.

\bibitem[{Brette et~al.(2007)Brette, Rudolph, Carnevale, Hines, Beeman, Bower,
  Diesmann, Morrison, Goodman, Harris~Jr et~al.}]{brette2007simulation}
Brette R, Rudolph M, Carnevale T, Hines M, Beeman D, Bower JM, Diesmann M,
  Morrison A, Goodman PH, Harris~Jr FC et~al. (2007) Simulation of networks of
  spiking neurons: a review of tools and strategies.
\newblock \emph{Journal of computational neuroscience} 23(3): 349--398.

\bibitem[{Calotoiu et~al.(2013)Calotoiu, Hoefler, Poke and
  Wolf}]{calotoiu_ea:2013:modeling}
Calotoiu A, Hoefler T, Poke M and Wolf F (2013) Using automated performance
  modeling to find scalability bugs in complex codes.
\newblock In: \emph{Proc. of the ACM/IEEE Conference on Supercomputing (SC13),
  Denver, CO, USA}. ACM, pp. 1--12.
\newblock \doi{10.1145/2503210.2503277}.

\bibitem[{Carnevale and Hines(2006)}]{carnevale2006neuron}
Carnevale NT and Hines ML (2006) \emph{The NEURON book}.
\newblock Cambridge University Press.

\bibitem[{Cremonesi et~al.(2019)}]{BBP-ECM-RA}
Cremonesi F et~al. (2019) Reproducibility appendix for paper on modeling {B}lue
  {B}rain {P}roject kernels with {ECM}.
\newblock
  \urlprefix\url{https://github.com/RRZE-HPC/BBP-ECM-RA/releases/tag/2019-01-16}.

\bibitem[{Fog(2017)}]{fog2017instruction}
Fog A (2017) Instruction tables: Lists of instruction latencies, throughputs
  and micro-operation breakdowns for intel, amd and via cpus. technical
  university of denmark.

\bibitem[{Gerstner et~al.(2014)Gerstner, Kistler, Naud and
  Paninski}]{gerstner2014neuronal}
Gerstner W, Kistler WM, Naud R and Paninski L (2014) \emph{Neuronal dynamics:
  From single neurons to networks and models of cognition}.
\newblock Cambridge University Press.

\bibitem[{Gruber et~al.(2018)}]{likwidweb}
Gruber T et~al. (2018) {LIKWID}: A multicore performance tool suite.
\newblock \urlprefix\url{http://tiny.cc/LIKWID}.

\bibitem[{Hager et~al.(2018)Hager, Eitzinger, Hornich, Cremonesi, Alappat,
  R{\"o}hl and Wellein}]{hager2018applying}
Hager G, Eitzinger J, Hornich J, Cremonesi F, Alappat CL, R{\"o}hl T and
  Wellein G (2018) Applying the execution-cache-memory model: Current state of
  practice
  \urlprefix\url{https://sc18.supercomputing.org/presentation/?id=post152\&sess=sess322}.
\newblock Poster at Supercomputing 2018.

\bibitem[{Hammer et~al.(2017)Hammer, Eitzinger, Hager and
  Wellein}]{hammer2017kerncraft}
Hammer J, Eitzinger J, Hager G and Wellein G (2017) Kerncraft: a tool for
  analytic performance modeling of loop kernels.
\newblock In: \emph{Tools for High Performance Computing 2016}. Springer, pp.
  1--22.

\bibitem[{Hines(1984)}]{hines1984efficient}
Hines M (1984) Efficient computation of branched nerve equations.
\newblock \emph{International journal of bio-medical computing} 15(1): 69--76.

\bibitem[{Hines et~al.(2011)Hines, Kumar and
  Sch{\"u}rmann}]{hines2011comparison}
Hines M, Kumar S and Sch{\"u}rmann F (2011) Comparison of neuronal spike
  exchange methods on a blue gene/p supercomputer.
\newblock \emph{Frontiers in computational neuroscience} 5: 49.

\bibitem[{Hines and Carnevale(2000)}]{hines2000expanding}
Hines ML and Carnevale NT (2000) Expanding neuron's repertoire of mechanisms
  with nmodl.
\newblock \emph{Neural computation} 12(5): 995--1007.

\bibitem[{Hofmann et~al.(2018)Hofmann, Hager and Fey}]{hofmann2018accuracy}
Hofmann J, Hager G and Fey D (2018) On the accuracy and usefulness of analytic
  energy models for contemporary multicore processors.
\newblock In: Yokota R, Weiland M, Keyes D and Trinitis C (eds.)
  \emph{International Conference on High Performance Computing}. Cham: Springer
  International Publishing, pp. 22--43.

\bibitem[{Hofmann et~al.(2017)Hofmann, Hager, Wellein and Fey}]{Hofmann:2017}
Hofmann J, Hager G, Wellein G and Fey D (2017) An analysis of core- and
  chip-level architectural features in four generations of {I}ntel server
  processors.
\newblock In: Kunkel JM, Yokota R, Balaji P and Keyes D (eds.)
  \emph{International Conference on High Performance Computing}. Cham: Springer
  International Publishing, pp. 294--314.

\bibitem[{Intel(2017)}]{IACA}
Intel (2017) {I}ntel {A}rchitecture {C}ode {A}nalyzer.
\newblock
  \urlprefix\url{https://software.intel.com/en-us/articles/intel-architecture-code-analyzer}.

\bibitem[{Intel(2018)}]{ia32opt:2018}
Intel (2018) {Intel 64 and IA-32 Architectures Optimization Reference Manual}.
\newblock
  \urlprefix\url{http://www.intel.com/content/dam/www/public/us/en/documents/manuals/64-ia-32-architectures-optimization-manual.pdf}.

\bibitem[{Ippen et~al.(2017)Ippen, Eppler, Plesser and
  Diesmann}]{ippen2017constructing}
Ippen T, Eppler JM, Plesser HE and Diesmann M (2017) Constructing neuronal
  network models in massively parallel environments.
\newblock \emph{Frontiers in neuroinformatics} 11: 30.

\bibitem[{Izhikevich and Edelman(2008)}]{izhikevich2008large}
Izhikevich EM and Edelman GM (2008) Large-scale model of mammalian
  thalamocortical systems.
\newblock \emph{Proceedings of the national academy of sciences} 105(9):
  3593--3598.

\bibitem[{Jordan et~al.(2018)Jordan, Ippen, Helias, Kitayama, Sato, Igarashi,
  Diesmann and Kunkel}]{jordan2018extremely}
Jordan J, Ippen T, Helias M, Kitayama I, Sato M, Igarashi J, Diesmann M and
  Kunkel S (2018) Extremely scalable spiking neuronal network simulation code:
  from laptops to exascale computers.
\newblock \emph{Frontiers in neuroinformatics} 12: 2.

\bibitem[{Kozloski and Wagner(2011)}]{kozloski2011ultrascalable}
Kozloski J and Wagner J (2011) An ultrascalable solution to large-scale neural
  tissue simulation.
\newblock \emph{Frontiers in neuroinformatics} 5: 15.

\bibitem[{Kumbhar et~al.(2016)Kumbhar, Hines, Ovcharenko, Mallon, King, Sainz,
  Sch{\"u}rmann and Delalondre}]{kumbhar2016leveraging}
Kumbhar P, Hines M, Ovcharenko A, Mallon DA, King J, Sainz F, Sch{\"u}rmann F
  and Delalondre F (2016) Leveraging a cluster-booster architecture for
  brain-scale simulations.
\newblock In: \emph{International Conference on High Performance Computing}.
  Springer, pp. 363--380.

\bibitem[{Laukemann et~al.(2018)Laukemann, Hammer, Hofmann, Hager and
  Wellein}]{Laukemann:2018}
Laukemann J, Hammer J, Hofmann J, Hager G and Wellein G (2018) Automated
  instruction stream throughput prediction for {Intel} and {AMD}
  microarchitectures.
\newblock \emph{CoRR} abs/1809.00912.
\newblock \urlprefix\url{http://arxiv.org/abs/1809.00912}.
\newblock Accepted for publication.

\bibitem[{Markram et~al.(2015)Markram, Muller, Ramaswamy, Reimann, Abdellah,
  Sanchez, Ailamaki, Alonso-Nanclares, Antille, Arsever
  et~al.}]{markram2015reconstruction}
Markram H, Muller E, Ramaswamy S, Reimann MW, Abdellah M, Sanchez CA, Ailamaki
  A, Alonso-Nanclares L, Antille N, Arsever S et~al. (2015) Reconstruction and
  simulation of neocortical microcircuitry.
\newblock \emph{Cell} 163(2): 456--492.

\bibitem[{McCalpin(1995)}]{McCalpin:1995}
McCalpin JD (1995) Memory bandwidth and machine balance in current high
  performance computers.
\newblock \emph{IEEE Computer Society Technical Committee on Computer
  Architecture (TCCA) Newsletter} : 19--25.

\bibitem[{NEC(2018)}]{tsubasa}
NEC (2018) {NEC} {SX}-{A}urora {TSUBASA} -- {V}ector {E}ngine.
\newblock
  \urlprefix\url{https://www.nec.com/en/global/solutions/hpc/sx/vector_engine.html}.

\bibitem[{Peyser and Schenck(2015)}]{peyser2015nest}
Peyser A and Schenck W (2015) The nest neuronal network simulator: Performance
  optimization techniques for high performance computing platforms.
\newblock In: \emph{Society for Neuroscience Annual Meeting}, FZJ-2015-06261.
  J{\"u}lich Supercomputing Center.

\bibitem[{Potjans and Diesmann(2012)}]{potjans2012cell}
Potjans TC and Diesmann M (2012) The cell-type specific cortical microcircuit:
  relating structure and activity in a full-scale spiking network model.
\newblock \emph{Cerebral cortex} 24(3): 785--806.

\bibitem[{Ramaswamy et~al.(2015)Ramaswamy, Courcol, Abdellah, Adaszewski,
  Antille, Arsever, Atenekeng, Bilgili, Brukau and
  Chalimourda}]{ramaswamy2015neocortical}
Ramaswamy S, Courcol JD, Abdellah M, Adaszewski SR, Antille N, Arsever S,
  Atenekeng G, Bilgili A, Brukau Y and Chalimourda Aea (2015) The neocortical
  microcircuit collaboration portal: a resource for rat somatosensory cortex.
\newblock \emph{Frontiers in neural circuits} 9: 44.
\newblock \urlprefix\url{https://bbp.epfl.ch/nmc-portal/downloads}.

\bibitem[{Schuecker et~al.(2017)Schuecker, Schmidt, van Albada, Diesmann and
  Helias}]{schuecker2017fundamental}
Schuecker J, Schmidt M, van Albada SJ, Diesmann M and Helias M (2017)
  Fundamental activity constraints lead to specific interpretations of the
  connectome.
\newblock \emph{PLoS computational biology} 13(2): e1005179.

\bibitem[{Stengel et~al.(2015)Stengel, Treibig, Hager and Wellein}]{sthw15}
Stengel H, Treibig J, Hager G and Wellein G (2015) Quantifying performance
  bottlenecks of stencil computations using the {E}xecution-{C}ache-{M}emory
  model.
\newblock In: \emph{Proceedings of the 29th ACM International Conference on
  Supercomputing}, ICS '15. New York, NY, USA: ACM.
\newblock \doi{10.1145/2751205.2751240}.
\newblock \urlprefix\url{http://doi.acm.org/10.1145/2751205.2751240}.

\bibitem[{Thomas(1949)}]{thomas1949elliptic}
Thomas LH (1949) Elliptic problems in linear difference equations over a
  network.
\newblock \emph{Watson Sci. Comput. Lab. Rept., Columbia University, New York}
  1.

\bibitem[{Tikidji-Hamburyan et~al.(2017)Tikidji-Hamburyan, Narayana, Bozkus and
  El-Ghazawi}]{tikidji2017software}
Tikidji-Hamburyan RA, Narayana V, Bozkus Z and El-Ghazawi TA (2017) Software
  for brain network simulations: a comparative study.
\newblock \emph{Frontiers in neuroinformatics} 11: 46.

\bibitem[{Treibig and Hager(2010)}]{treibig2010introducing}
Treibig J and Hager G (2010) Introducing a performance model for
  bandwidth-limited loop kernels.
\newblock In: \emph{Parallel Processing and Applied Mathematics}. Springer, pp.
  615--624.

\bibitem[{Treibig et~al.(2010)Treibig, Hager and Wellein}]{psti}
Treibig J, Hager G and Wellein G (2010) Likwid: A lightweight
  performance-oriented tool suite for x86 multicore environments.
\newblock In: \emph{Proceedings of PSTI2010, the First International Workshop
  on Parallel Software Tools and Tool Infrastructures}. San Diego CA.

\bibitem[{Williams et~al.(2009)Williams, Waterman and
  Patterson}]{roofline:2009}
Williams S, Waterman A and Patterson D (2009) Roof{}line: An insightful visual
  performance model for multicore architectures.
\newblock \emph{Commun. ACM} 52(4): 65--76.
\newblock \doi{10.1145/1498765.1498785}.
\newblock \urlprefix\url{http://doi.acm.org/10.1145/1498765.1498785}.

\bibitem[{Zenke and Gerstner(2014)}]{zenke2014limits}
Zenke F and Gerstner W (2014) Limits to high-speed simulations of spiking
  neural networks using general-purpose computers.
\newblock \emph{Frontiers in neuroinformatics} 8: 76.

\end{thebibliography}

\end{document}